\documentclass{elsart1p}

\usepackage{amssymb}

\usepackage{amsmath} 
\usepackage{mathrsfs} 
\usepackage{txfonts} 
\usepackage{stmaryrd} 
\usepackage{graphicx}        
\usepackage{amssymb}  

\begin{document}

\begin{frontmatter}



\title{Physical limits of inference}


\author{David H. Wolpert}
\address{MS 269-1, NASA Ames Research Center, Moffett Field, CA 94035,
USA}

\ead{dhw@ptolemy.arc.nasa.gov}
\ead[url]{ti.arc.nasa.gov/people/dhw}

\begin{abstract}
I show that physical devices that perform observation, prediction, or
recollection share an underlying mathematical structure. I call
devices with that structure ``inference devices''.  I present a set of
existence and impossibility results concerning inference
devices. These results hold independent of the precise physical laws
governing our universe. In a limited sense, the impossibility results
establish that Laplace was wrong to claim that even in a classical,
non-chaotic universe the future can be unerringly predicted, given
sufficient knowledge of the present. Alternatively, these
impossibility results can be viewed as a non-quantum mechanical
``uncertainty principle''. Next I explore the close connections
between the mathematics of inference devices and of Turing Machines.
In particular, the impossibility results for inference devices are
similar to the Halting theorem for TM's. Furthermore, one can define
an analog of Universal TM's (UTM's) for inference devices. I call
those analogs ``strong inference devices''. I use strong inference
devices to define the ``inference complexity'' of an inference task,
which is the analog of the Kolmogorov complexity of computing a
string. However no universe can contain more than one strong inference
device. So whereas the Kolmogorov complexity of a string is arbitrary
up to specification of the UTM, there is no such arbitrariness in the
inference complexity of an inference task. I end by discussing the
philosophical implications of these results, e.g., for whether the
universe ``is'' a computer.
\end{abstract}

\end{frontmatter}

\nocite{lewi86,geha86,kant90,berg90,dado91,gell96,toll00,ruoh97,hodg97,schm02}

\section{Introduction}

Some of the most fruitful investigations of the foundations of physics
began by identifying a set of features that are present in all
physical realizations of a particular type of information
processing. The next step in these investigations was to abstract and
formalize those shared features.  Once that was done, one could
explore the mathematical properties of those features, and thereby
analyze some aspects of the relationship between physics and
information processing.  Examples of such investigations include the
many decades of work on the relationship between physics and
computation~\cite{wolp92b,lloy06,lloy00,zurk84,land61,land88,moor90,pori82,frto82,feyn86,benn73,benn82,benn87b,benn82,aaro05},
the work on observation that started with Everett's seminal
paper~\cite{ever57}, and more recent work that considers what possible
forms physical reality might
have~\cite{wolp01,smol02,agte05,carr07,barb99,coko06,wolf02,tegm07,mcca06,davi07,schm97}.

In this spirit, here we first present archetypal examples of physical
devices that perform observation, of physical devices that perform
prediction, and of physical devices that perform recollection. We then
identify a set of features common to those examples.  This is our
first contribution, that such physical devices share those features.

Next we formalize those features, defining any device possessing them
to be an ``inference device''.  To do this requires our second
contribution: a formalization of the concept of semantic information
content.{\footnote{In contrast to the concept of syntactic information
content, whose formalization by Shannon is the basis of conventional
information theory~\cite{coth91}.}}  Loosely speaking, we define the
semantic information content of a variable $s$ concerning a variable
$r$ to be what an external scientist can infer about what the value of
$r$ is in their particular universe by knowing the state of $s$. Note
the central role in this definition of the scientist external to the
device. As discussed below, in the context of using inference devices
for observation, this central role of the external scientist is in
some ways more consistent with Wigner's view of the observation
process than with the many-worlds view of that process.

For the remainder of the paper we develop the theory of inference
devices, thereby analyzing numerous aspects of the relationship
between physics and information processing. Our goal in this endeavor
is to illustrate the breadth of the theory of inference devices; an
exhaustive analysis of any one aspect of that theory is beyond what
can fit into this single paper.

A recurring theme in our analysis of inference devices is their
relationship with Turing Machines (TM's). In particular, there are
impossibility results for inference devices that are similar to the
Halting theorem for TM's. Furthermore, one can define an analog of
Universal TM's (UTM's) for inference devices. We call those analogs
``strong inference devices''.

A central result of this paper is how to use strong inference devices
to define the ``inference complexity'' of an inference task, which is
the analog of the Kolmogorov complexity of computing a string.  A
task-independent bound is derived on how much the inference complexity
of an inference task can differ for two different inference
devices. This is analogous to the ``encoding'' bound governing how
much the Kolmogorov complexity of a string can differ between two
UTM's used to compute that string.  However no universe can contain
more than one strong inference device. So whereas the Kolmogorov
complexity of a string is arbitrary up to specification of the UTM,
there is no such arbitrariness in the inference complexity of an
inference task.

After presenting inference complexity, we informally discuss the
philosophical implications of all of our results to that point. In
particular, we discuss what it might mean for the universe to ``be'' a
computer. We also show how much of philosophy can be reduced to
constraint satisfaction problems, potentially involving
infinite-dimensional spaces.  We follow this discussion by deriving
some graph-theoretic properties governing the possible inference
relationships among any set of multiple inference devices in the same
universe.

Our next contribution is an extension of the inference devices
framework to include physical devices that are used for
control. Associated impossibility results provide fundamental limits
on the capabilities of physical control systems.  After this we
present an extension of the framework to probabilistic inference
devices. Of all the results in this paper, it is the impossibility
results concerning probabilistic inference devices that are the most
similar to quantum mechanical impossibility results. We end by presenting
an extension of the framework that clarifies its relation with
semantic information. 

The crucial property underlying our results is that inference devices
are embodied in the very physical system (namely the universe) about
which they are making inferences.  This embedding property and its
consequences have nothing to do with the precise laws governing the
underlying universe. In particular, those consequences do not involve
chaotic dynamics as in~\cite{moor90,pori82}, nor quantum mechanical
indeterminism. Similarly, they apply independent of the values of any
physical constants (in contrast, for example, to the work
in~\cite{lloy06}), and more generally apply to every universe in a
multiverse. Nor do the results presume limitations on where in the
Chomsky hierarchy an inference device lies. So for example they would
apply to oracles, if there can be oracles in our universe. In the
limited sense of our impossibility results, Laplace was wrong to claim
that even in a classical, non-chaotic universe the future can be
unerringly predicted, given sufficient knowledge of the
present~\cite{lapl85}. Alternatively, these impossibility results
can be viewed as a non-quantum mechanical ``uncertainty principle''.

All non-trivial proofs are in App. A. An earlier analysis addressing
some of the issues considered in this paper can be found
in~\cite{wolp01}.

\subsection{Notation}
\label{sec:notation}

We will take the set of binary numbers $\mathbb{B}$ to equal $\{-1,
1\}$, so that logical negation is indicated by the minus sign. We will
also take $\Theta$ to be the Heaviside theta function that equals 1 if
its argument is non-negative, 0 otherwise. $\mathbb{N}$ is the
natural numbers, $1, 2, \ldots$. For any function $\Gamma$ with domain
$U$, we will write the image of $U$ under $\Gamma$ as $\Gamma(U)$. For
any function $\Gamma$ with domain $U$ that we will consider, we
implicitly assume that $\Gamma(U)$ contains at least two distinct
elements. For any (potentially infinite) set $W$, $|W|$ is the
cardinality of $W$. For any real number $a \in {\mathbb{R}}$, $\lceil
a$ is the smallest integer greater than or equal to $a$. Given two
functions $\Gamma_1$ and $\Gamma_2$ with the same domain $U$, we write
$\Gamma_1 \otimes \Gamma_2$ for the function with domain $U$ obeying
$u \in U : \rightarrow (\Gamma_1(u), \Gamma_2(u))$, and with some
abuse of terminology refer to this as the ``product'' of
$\Gamma_1$ and $\Gamma_2$.

Given a function $\Gamma$ with domain $U$, we say that the partition
{\bf{induced}} by $\Gamma$ is the family of subsets
$\{\Gamma^{-1}(\gamma) : \gamma \in \Gamma(U)\}$. Intuitively, it is
the family of subsets of $U$ each of which consists of all elements
having the same image under $\Gamma$.  We will say that a partition
$A$ over a space $U$ is a {\bf{fine-graining}} of a partition $B$ over
$U$ (or equivalently that $B$ is a coarse-graining of $A$) iff every
$a \in A$ is a subset of some $b \in B$. Two partitions $A$ and $B$
are fine-grainings of each other iff $A = B$. Say a partition $A$ is
finite and a fine-graining of a partition $B$. Then $|A| = |B|$ iff $A
= B$. 

Given a probability measure, the mutual information between two
associated random variables $a, b$ conditioned on event $c$ is written
${\mathbb{M}}(a, b \mid c)$. The Shannon entropy of random variable
$a$ is ${\mathbb{H}}(a)$.

\section{Archetypal examples}
\label{sec:examples}

We now illustrate that many (if not all) physical realizations of the
processes of observation, prediction, and memory share a certain
mathematical structure. We do this by semi-formally describing each of
those processes, one after the other. Each such description uses
language that is purposely very similar to the other descriptions. It
is that very similarity of language that demonstrates that the same
mathematical structure arises as part of each of the processes. In the
following sections of this paper we will formalize that mathematical
structure, and then present our formal results concerning
it.{\footnote{Some might quibble that one or another of the these
examples should involve additional structure, that what is presented
in that example does not fully capture the physical processes it
claims to describe.  (See App. B.) The important point is that the
structure presented in these examples is always found in real-world
instances of the associated physical processes.  Whether or not there
is additional structure that ``should'' be assumed is not relevant.
The structure that is assumed in the examples is sufficient to
establish our formal results.}}

If the reader becomes convinced of this shared mathematical structure
before reading through all the examples, (s)he is encouraged to skip
to the next section.  It is in that section that we formalize the
shared mathematical structure, as an ``inference device''.

In all of the examples in this section, $U$ is the space of all
worldlines of the entire universe that are consistent with the laws of
physics (whatever they may be), and $u$ indicates an element of
$U$.{\footnote{For expository simplicity we use the language of
non-quantum mechanical systems in this paper. However most of what
follows holds just as well for a quantum-mechanical universe, if we
interpret quantum mechanics appropriately.}}

$ $

\noindent {\bf{Example 1:}} We start by describing 
a physical system that is a general-purpose observation device,
capable of observing different aspects of the universe. Let $S$ be
some particular variable concerning the universe whose value at some
time $t_2$ we want our device to observe. If the universe's worldline
is $u$, then the value of $S$ at $t_2$ is given by some function of
$u$ (e.g., it could be given by a component of $u$). Write that
function as $\Gamma$; $S(t_2) = \Gamma(u)$.

The observation device consists of two parts: an observation
apparatus, and a scientist who uses (and interprets) that
apparatus. To make our observation, the scientist must first configure
the observation apparatus to be in some appropriate state at some time
$t_1 < t_2$. (The idea is that by changing how the observation
apparatus is configured the scientist can change what aspect of the
universe he observes.)  That configuration of the observation
apparatus at $t_1$ is also given by a function of the entire
universe's worldline $u$, since the observation apparatus exists in
the universe. Write that function as $\chi$, with range $\chi(U)$.

The goals is that if the apparatus has been properly configured, then
sometime after $t_1$ it couples with $S$ in such a way that at some
time $t_3 > t_2$, the output display of the observation apparatus
accurately reflects $S(t_2)$. Again, that output display exists in the
universe. So its state at $t_3$ is a function of $u$; write that
function as $\zeta$.

The scientist reads the output of the apparatus and interprets that
output as this attempted observation of $S(t_2)$. It is this
interpretation that imbues that output with semantic
information. Without such interpretation the output is just a
meaningless (!) pattern, one that happens to be physically coupled
with the variable being observed. (As an extreme example of such
meaningless coupling, if a tree falls in a forest, but the video that
recorded the fall is encrypted in a way that the scientist cannot
undo, then the scientist does not ``observe'' that the tree fell by
watching the video .)

To formalize what such interpretation means, we must define ``semantic
information''. As mentioned above, we want the semantic information of
a variable $s$ concerning a variable $r$ to be what an external
scientist can infer about $r$ by knowing the state of $s$.  In the
current example this means we require that the scientist can ask
questions of the sort, ``Does $S(t_2) = K$?'' at $t_3$, and that
$\zeta(u)$ provides the scientist with (possibly erroneous) answers to
such questions. As an example, say that $\zeta(u)$ is a display
presenting integers from $0$ to $1000$, inclusive, with a special
'error' symbol for integers outside of that range. Since the scientist
interprets the value on that display at $t_3$ as the outcome of the
observation of $S(t_2)$, by looking at the display at $t_3$ the
scientist is provided with (possibly erroneous) answers to the
question ``Does $S(t_2) = K$?'' for all $1001$ values of $K$ that can
be on the display.

To make this more precise, first note that any question like ``Does
$S(t_2) = K$?'' can either be answered 'yes' or 'no', and therefore is
a binary function of $u$. For every $K$, write this associated binary
function of $u$ as $q_K$; $\forall K, \forall u \in U, q_K(u) = 1$ if
$S(t_2) = \Gamma(u) = K$, and it equals -1 otherwise. Next, note that
the brain of the scientist exists in the universe. So which (if any)
of a set of such possible binary questions concerning the universe the
scientist is asking at $t_3$ is also a function of $u$. We write that
function as $Q$. In particular, we presume that any question $q_K$ is
one of the elements in the range of $Q$, i.e., it is one of the
questions that (depending on the state of the scientist's brain then)
the scientist might be asking at $t_3$.

Now for any particular question $q_K$ the scientist might be asking at
$t_3$, the answer that the scientist provides by interpreting the
apparatus' output is a bit. The value of that bit is specified by the
state of the scientist's brain at $t_3$. (The premise being that the
state of the scientist's brain was affected by the scientist's reading
and then interpreting the apparatus' output.) So again, since the
scientist's brain exists in the universe, the value of that answer bit
is a function of $u$.  We write that function as $Y$.

It is the combination of $Q$ and $Y$ that comprise the scientist's
``interpretation'' of $\zeta$, and thereby imbue any particular
$\zeta(u)$ with semantic content. $Q(u)$ specifies a question $q_K$.
$\zeta(u)$ then causes $Y(u)$ to have some associated value. We take
that value to be (the scientist's interpretation of) the apparatus'
answer to the question of whether $q_K(u) = 1$ or $q_K(u) = -1$ (i.e.,
of whether $S(t_2) = K$). Combining, $\zeta(u)$ causes $Y(u)$ to have
a value that we take to be (the scientist's interpration of) the
apparatus' answer to whether $[Q(u)](u) = 1$ or $[Q(u)](u) = -1$.

This scenario provides a set of requirements for what it means for the
combination of the observation apparatus and the scientist using that
apparatus to be able to successfully observe the state of $S$ at
$t_2$: First, we require that the scientist can configure the
apparatus in such a way that its output at $t_3$ gives $\Gamma(u)$. We
also require that the scientist can read and interpret that
output. This means at a minimum that for any question of the form
``Does $\Gamma(u) = K$?''  the scientist can both ask that question at
$t_3$ and interpret $\zeta(u)$ to accurately answer it.

To make this fully formal, we introduce a set of binary functions with
domain $\Gamma(U)$: $\forall K, f_K : \gamma
\rightarrow 1$ iff $\gamma = K$. Note that we  have one such 
function for every $K \in \Gamma(U)$. Our requirement for successful
observation is that the observation apparatus can be configured so
that, for any $f_K$, if the scientist were to consider an associated
binary question at $t_3$ and interpret $\zeta(u)$ to answer the
question, then the scientist's answer would necessarily equal
$f_K(\Gamma(u))$. In other words, there is a value $c \in \chi(U)$
such that for any $K \in \Gamma(U)$, there is an associated $q_K \in
Q(U)$ such that the combination of $\chi(u) = c$ and $Q(u) = q_K$
implies that $Y(u) = f_K(\Gamma(u))$.

Intuitively, for the scientist to use the apparatus to ``observe
$S(t_2)$'' only means the scientist must configure the apparatus
appropriately; the scientist must force the universe to have a
worldline $u$ such that $\chi(u) = c$, and that must in turn cause
$\zeta(u)$ to accurately give $\Gamma(u)$. In particular, to ``observe
$S(t_2)$'' does not require that the scientist impose any particular
value on $Q(u)$. Rather $Q$'s role is to provide a way to interpret
$\zeta(u)$. The only requirement made of $Q$ is that {\it{if}} the
scientist were to ask a question like ``Does $S(t_2)$ equal $K$?'',
then $Q(u)$ --- determined by the state of the scientist's brain at
$t_3$ --- would equal that question, and the scientist's answer $Y(u)$
would be appropriately set by $\zeta(u)$. It is by using $Q$ this way
that we formalize the notion that $\zeta(u)$ conveys information to
the scientist concerning $S(t_2)$.  The ``observation is successful''
if for any such question the scientist $might$ pose (as reflected in
$Q(u)$), their associated answer (as reflected in $Y(u)$) properly
matches the state of $S$ at $t_2$. 

We can motivate this use of $Q$ in a less nuanced, more direct
way. Consider a scenario where the scientist can$not$ both pose all
binary-valued questions $f_K$ concerning $S(t_2)$ and correctly answer
them using the apparatus output, $\zeta(u)$. It would seem hard to
justify the view that in this scenario the combination of the
scientist with the apparatus makes a ``successful observation''
concerning $S(t_2)$.

Note that by defining an observation device as the combination of an
observation apparatus with the external scientist who is using that
apparatus, we are in a certain sense arriving at a Wignerian approach
to observation. In contrast to a more straight-forward many-worlds
approach, we require that the state of the observation apparatus not
just be correlated with the variable being observed, but in fact
contain semantic information concerning the variable being
observed. This makes the external scientist using the observation
apparatus crucial in our approach, in contrast to the case with the
many-worlds approach.

$ $

\noindent {\bf{Example 2:}} This example is a slight variant of
Ex. 1. In this variant, there is no scientist, just ``inanimate''
pieces of hardware. 

We change the apparatus of Ex. 1 slightly. First, we make the output
$\zeta$ be binary-valued. We also change the configuration function
$\chi$, so that in addition to its previous duties, it also specifies
a question of the form, ``Does $\Gamma(u)$ equal $K$?''. Then observation is
successful if for any $K \in
\Gamma(U)$, the apparatus can be configured appropriately, so that its output correctly
answers the question of whether $S(t_2)$ equals $K$. In other words,
observation is successful if for any $K \in \Gamma(U)$ there is an
associated $c \in \chi(U)$ such that having $\chi(u) = c$ implies that
$Y(u) = f_K(\Gamma(u))$.

$ $

\noindent {\bf{Example 3:}} We now describe
a physical system that is a general-purpose prediction device, capable
of correctly predicting different aspects of the universe's
future. Let $S$ be some particular variable concerning the universe
whose value at some time $t_2$ we want our device to predict. If the
universe's worldline is $u$, then the value of $S$ at $t_2$ is given
by some function of $u$ which we write as $\Gamma$; $S(t_2) =
\Gamma(u)$.

The prediction device consists of two parts, a physical computer, and
a scientist who programs that computer to make the prediction and
interprets the computer's output as that prediction. To ``program the
computer'' means that the scientist initializes it at some time $t_1 <
t_2$ to contain some information concerning the state of the universe
and to run a simulation of the dynamics of the universe that uses that
information. Accordingly, to ``program the computer'' to perform the
prediction means making it be in some appropriate state at $t_1$. (The
idea is that by changing how the computer is programmed, the
scientist can change what aspect of the universe the computer
predicts.)  That initialization of the computer is also given by a
function of the entire universe's worldline $u$, since the computer
exists in the universe. Write that function as $\chi$, with range
$\chi(U)$.

The hope is that if the computer is properly programmed at $t_1$, then
it runs a simulation concerning the evolution of the universe that
completes at some time $t_3 > t_1$, and at that time displays a
correct prediction of $S(t_2)$ on its output. (In general we would
like to also have $t_3 < t_2$, so that the simulation completes before
the event being predicted actually occurs, but we don't require that.)
Again, that output display exists in the universe. So its state at
$t_3$ is a function of $u$; write that function as $\zeta$.

The scientist reads the output of the computer and interprets it as
this attempted prediction of $S(t_2)$, thereby imbuing that output
with semantic meaning. More precisely, for the value $\zeta(u)$ to
convey information to the scientist at $t_3$, we require that the
scientist can ask questions of the sort, ``Does $S(t_2) = K$?'' at
$t_3$, and that $\zeta(u)$ provides the scientist with (possibly
erroneous) answers to such questions.

As in Ex. 1, to make this more formal, we note that any question like
``Does $S(t_2) = K$?'' is a binary function of $u$, of the sort $q_K$
presented in Ex. 1. Also as in Ex. 1, the brain of the scientist
exists in the universe. So which (if any) of a set of possible
questions concerning the universe the scientist is asking at $t_3$ is
also a function of $u$, which we again write as $Q$. Also as in Ex. 1,
the answer of the scientist to any such question is a bit that the
scientist generates by interpreting $\zeta(u)$. Since that answer is
given by the state of the scientist's brain at $t_3$, it is a function
of $u$, which as before we write as $Y$.

So for the combination of the computer and the scientist using that
computer to be able to successfully predict the state of $S$ at $t_2$
means two things: First, we require that the scientist can program the
computer in such a way that its output at $t_3$ gives $\Gamma(u)$. We
also require that the scientist can read and interpret that
output. More precisely, our requirement for successful prediction is
that the computer can be programmed so that, for any $f_K$, if the
scientist were to consider an associated binary question at $t_3$ and
interpret $\zeta(u)$ to answer the question, then the scientist's
answer would necessarily equal $f_K(\Gamma(u))$.  In other words,
there is a value $c \in \chi(U)$ such that for any $K \in \Gamma(U)$,
there is an associated $q_K \in Q(U)$ such that the combination of
$\chi(u) = c$ and $Q(u) = q_K$ implies that $Y(u) = f_K(\Gamma(u))$.

Just as in Ex. 1, for the scientist to use the apparatus to ``predict
$S(t_2)$'' only means the scientist must program the computer
appropriately; the scientist must force the universe to have a
worldline $u$ such that $\chi(u) = c$, and that must in turn cause
$\zeta(u)$ to accurately give $\Gamma(u)$. In particular, to ``predict
$S(t_2)$'' does not require that the scientist impose any particular
value on $Q(u)$. As before,  $Q$'s role is to provide a way to interpret
$\zeta(u)$. 

Note that the ``computer'' in this example is defined in terms of what
it does, not in terms of how it does it. This allows our formalization
of prediction to avoid all issues of where exactly in the Chomsky
hierarchy some particular physical computer might lie.

$ $

Nothing in the formalizations ending Ex.'s 1 - 3 relies on the precise
choices of time-ordering imposed on the values $t_1, t_2, t_3,
t_4$. Those formalizations only concern relations between functions
$\Gamma, f_k, Q, \zeta$ and $Y$, each having the entire worldline
across all time as its domain. This fact means that the same sort of
formalization can be applied to ``retrodiction'', as elaborated in the
following example.

$ $

\noindent {\bf{Example 4:}} Say we have a system that we want to serve
as a general-purpose recording and recollection device, capable of
correctly recording different aspects of the universe and recalling
them at a later time. Let $S$ be some particular variable concerning
the universe whose value at some time $t_2$ we want our device to
record. If the universe's worldline is $u$, then the value of $S$ at
$t_2$ is given by some function of $u$ which we write as the function 
$\Gamma$; $S(t_2) = \Gamma(u)$.

The recording device consists of two parts. The first is a physical
recording apparatus that records many characteristics of the
universe. The second is a scientist who queries that apparatus to see
what it has recorded concerning some particular characteristic of the
universe, and interprets the apparatus' response as that recording. To
``query the apparatus'' means that the scientist makes some variable
concerning the apparatus be in an appropriate state at some time $t_1
> t_2$. (The idea is that by changing how the apparatus is queried,
the scientist can change what aspect of the universe's past the
apparatus displays to the scientist.)  That state imposed on the
variable concerning the apparatus at $t_1$ is also given by a function
of the entire universe's worldline $u$, since the apparatus exists in
the universe. Write that function as $\chi$, with range $\chi(U)$.

The hope is that if the apparatus functions properly and is properly
queried at $t_1$, then it retrieves an accurate recording of $S(t_2)$,
and displays that recording on its output at some time $t_3 >
t_1$. Again, that output display of the apparatus exists in the
universe. So its state at $t_3$ is a function of $u$; write that
function as $\zeta$.

The scientist reads the output of the apparatus and interprets it as
this recording of $S(t_2)$, thereby imbuing that output
with semantic meaning. More precisely, for the value $\zeta(u)$ to
convey information to the scientist at $t_3$, we require that the
scientist can ask questions of the sort, ``Does $S(t_2) = K$?'' at
$t_3$, and that $\zeta(u)$ provides the scientist with (possibly
erroneous) answers to such questions.

As in Ex. 1, to make this more formal, we note that any such question
is a binary function of $u$, of the sort $q_K$ presented in
Ex. 1. Also as in Ex. 1, the brain of the scientist exists in the
universe. So which (if any) of a set of possible questions concerning
the universe the scientist is asking at $t_3$ is also a function of
$u$, which we again write as $Q$. Also as in Ex. 1, the answer of the
scientist to any such question is a bit that the scientist generates
by interpreting $\zeta(u)$. Since that answer is given by the state of
the scientist's brain at $t_3$, it is a function of $u$, which as
before we write as $Y$.

So for the combination of the apparatus and the scientist using that
apparatus to be able to successfully record and recall the state of
$S$ at $t_2$ means two things: First, we require that the scientist
can query the apparatus in such a way that its output at $t_3$ gives
$\Gamma(u)$. We also require that the scientist can read and interpret
that output. More precisely, our requirement for successful recording
and recollection is that the apparatus can be queried so that, for any
$f_K$, if the scientist were to consider an associated binary question
at $t_3$ and interpret $\zeta(u)$ to answer the question, then the
scientist's answer would necessarily equal $f_K(\Gamma(u))$.  In other
words, there is a value $c \in \chi(U)$ such that for any $K \in
\Gamma(U)$, there is an associated $q_K \in Q(U)$ such that the
combination of $\chi(u) = c$ and $Q(u) = q_K$ implies that $Y(u) =
f_K(\Gamma(u))$.

Just as in Ex. 1, for the scientist to use the apparatus to ``recall
$S(t_2)$'' only means the scientist must query the apparatus
appropriately; the scientist must force the universe to have a
worldline $u$ such that $\chi(u) = c$, and that must in turn cause
$\zeta(u)$ to accurately give $\Gamma(u)$. In particular, to ``recall
$S(t_2)$'' does not require that the scientist impose any particular
value on $Q(u)$. As before, $Q$'s role is to provide a way to
interpret $\zeta(u)$.

Note that nothing in this example specifies how the recording process
operates. This is just like how nothing in Ex. 1 specifies how the
observation apparatus couples with $S$, and how nothing in Ex. 3
specifies what simulation the computer runs.

See~\cite{wolp92a,wolp92b,barb99} for discussion about the crucial
role that recollection devices play in the psychological arrow of
time, and of the crucial dependence of such devices on the second law
of thermodynamics. As a result of their playing such a role, the
limitations on recollection devices derived below have direct
implications for the psychological and thermodynamic arrows of time.

$ $

Just as Ex. 2 varies Ex. 1 by removing the scientist, so Ex.'s 3 and 4
can be varied to remove the scientist.

\section{Basic concepts}

In this section we first formalize the mathematical structure that is
shared among Ex.'s 1-4 of Sec.~\ref{sec:examples}. In doing so we
substantially simplify that structure. After this formalization of the
shared structure in the examples we present some elementary results
concerning that structure.

\subsection{Inference devices}

\noindent {\bf{Definition 1}:} An {\bf{(inference) device}} over a set
$U$ is a pair of functions $(X, Y)$, both with domain $U$. $Y$ is
called the {\bf{conclusion}} function of the device, and is surjective
onto $\mathbb{B}$. $X$ is called the {\bf{setup}} function of the
device.

$ $

\noindent As an illustration, in all of Ex.'s 1-4, the setup function
is the composite function $(\chi, Q)$, and the conclusion function is
$Y$. The value of $X(u)$ can loosely be interpreted as how the device
is ``initialized / configured''.{\footnote{Care should be taken with
this interpretation though. For example, in Ex. 1, $\chi$ concerns the
state of $u$ at time $t_1$, and $Q$ concerns the state of $u$ at
$t_3$. So $X$ ``straddles multiple times''.}}  The value of $Y(u)$
should instead be viewed as all that the device predicts /observes /
recollects when it is done. {\it{A priori}}, we assume nothing about
how $X$ and $Y$ are related. Note that we do not require that the
compound map $(X, Y) : u \in U \rightarrow (X,Y)(u)$ be
surjective. There can be pairs of values $x
\in X(U)$, $y \in Y(U)$ that never arise for the same $u$.

Given some function $\Gamma$ with domain $U$ and some $\gamma \in
\Gamma(U)$, we are interested in setting up a device so that it is
assured of correctly answering whether $\Gamma(u) = \gamma$ for the
actual universe $u$.  Loosely speaking, we will formalize this with
the condition that $Y(u) = 1$ iff $\Gamma(u) =
\gamma$ for all $u$ that are consistent with some associated setup
value of the device, i.e., such that $X(u) = x$.  If this condition
holds, then setting up the device to have setup value $x$ guarantees
that the device will make the correct conclusion concerning whether
$\Gamma(u) = \gamma$. (Hence the terms ``setup function'' and
``conclusion function'' in Def. 1.)

Note that this desired relationship between $X$, $Y$ and $\Gamma$ can
hold even if $X(u) = x$ doesn't fix a unique value for $Y(u)$. Such
non-uniqueness is typical when the device is being used for
observation. Setting up a device to observe a variable outside of that
device restricts the set of possible universes; only those $u$ are
allowed that are consistent with the observation device being set up
that way to make the desired observation. But typically just setting
up an observation device to observe what value a variable has doesn't
uniquely fix the value of that variable.

In general we will want to predict / observe / recollect a function
$\Gamma$ that can take on more than two values. This is done by
appropriately choosing $X(u)$. As mentioned, $X(u)$ specifies what is
known about the outside world together with a simulation program (in
the case of computer-based prediction), or a specification of how to
set up an observation apparatus (in the case of observation), or a
specification of what to remember (in the case of a memory
device). But in addition, in all those cases $X(u)$ specifies one of
the possible values of $\Gamma(u)$ (i.e., it specifies a question of
the form ``Does $\Gamma(u) = \gamma$?''). We then view the device's
conclusion bit as saying whether $\Gamma(u)$ does / doesn't have that
specified value. So for example if our device is a computer being used
to predict the value of some variable concerning the state of the
world, then formally speaking, the setup of the computer specifies a
particular one of the possible values of that variable (in addition to
specifying other information like what simulation to run, what is
known about the outside world, etc.). Our hope is that the computer's
conclusion bit correctly answers whether the variable has that value
specified in how the computer is set up.

Intuitively, this amounts to using a unary representation of
$\Gamma(U)$.  To formalize this with minimal notation, we will use the
following shorthand:

$ $

\noindent {\bf{Definition 2:}} Let $A$ be a set having at least two
elements.  A {\bf{probe}} of $A$ is a mapping from $A$ onto
$\mathbb{B}$ that equals $1$ for one and only one argument $a \in A$.

$ $

\noindent So a probe of $A$ is a function that picks out a single one of $A$'s
possible values, i.e., it is a Kronecker delta function whose second
argument is fixed, and whose image value 0 is replaced by -1.

\subsection{Notation for inference devices}

We now have the tools to define what it means for an inference device
to successfully observe / predict / recall. Before presenting that
definition we introduce some useful notation.

Unless specified otherwise, a device written as ``$C_i$'' for any
integer $i$ is implicitly presumed to have domain $U$, with setup
function $X_i$ and conclusion function $Y_i$ (and similarly for no
subscript). Similarly, unless specified otherwise, expressions like
``min$_{x_i}$'' mean min$_{x_i \in X_i(U)}$.

We define a probe of a device to be a probe of the image of the
device's conclusion function. Given a function $\Gamma$ with domain
$U$ and a probe $f$ of $\Gamma(U)$, we write $f(\Gamma)$ as shorthand
for the function $u \in U \rightarrow f(\Gamma(u))$. We write $\pi(A)$
to indicate the set of all probes of a set $A$, and $\pi(\Gamma)$ to
indicate the set of functions over $U$, $\{f(\Gamma) : f \in
\pi(\Gamma(U))\}$. 

Probes are a shorthand way of posing queries concerning membership in a set (e.g., queries like ``is it true that $u \in Y^{-1}(y)$ for some particular value $y$?"). All such queries are binary-valued (which is why the range of probes is $\mathbb{B}$).  So couching the analysis in terms of probes essentially amounts to representing all associated spaces in terms of bits. This has the advantage that it allows us to avoid considering the ranges of any functions that arise in the analysis. In particular, it allows us to avoid concern
for whether one such range ``matches up'' with the domains and/or ranges
of other functions. For example, it allows us to avoid concern for such matching between the spaces defining two different inference devices when considering whether they infer each other.. (See~\cite{wolp01} for a more elaborate way of circumventing
the need of those ranges to match.)

Say we are given a set of functions over $U$, $\{D_1, d_1, D_2, d_2,
\ldots E_1, e_1, E_2, e_2, \ldots\}$. Then with some abuse of
terminology, we write ``$D_1 = d_1, D_2 = d_2, \ldots
\Rightarrow E_1 = e_1, E_2 = e_2, \ldots$'' as shorthand for
``$\exists \; u \in U$ such that $D_1(u) = d_1(u), D_2 = d_2, \ldots$,
and $\forall \; u \in U$ such that $D_1(u) = d_1(u), D_2 = d_2, \ldots$, it is
the case that $E_1(u) = e_1(u), e_2(u) = E_2(u), \ldots$''. We will
often abuse notation even further by allowing $d_1$ to be an element
of $D_1$'s range. In this case, ``$D_1 = d_1 \Rightarrow E_1 = e_1$''
is shorthand for``$\exists u \in U$ such that $D_1 = d_1$, and $\forall
\; u \in U$ such that $D_1(u) = d_1$, it is also the case that $E_1(u)
= e_1(u)$''.

\subsection{Weak inference}
\label{sec:def_weak}

We can now formalize inference as follows:

$ $

\noindent {\bf{Definition 3:}} A device $C$ {\bf{(weakly)
infers}} a function $\Gamma$ over ${U}$ iff $\forall f \in 
\pi(\Gamma)$, $\exists \; x$ such that $X = x \Rightarrow Y =
f({\Gamma})$.

$ $

\noindent So using the definitions in the previous subsection, $C$
weakly infers $\Gamma$ iff $\forall f \in 
\pi(\Gamma)$, $\exists \; x \in X(U)$ such that for all $u \in U$ for which
$X(u) = x$, $Y(u) = f({\Gamma}(u))$.

Recall our stipulation that all functions over $U$ take on at least
two values, and so in particular $\Gamma$ must. Therefore
$\pi(\Gamma)$ is non-empty. We will write $C > \Gamma$ if $C$ infers
$\Gamma$. Expanding our shorthand notation, $C > \Gamma$ means that
for all $\gamma \in \Gamma(U)$, $\exists x \in X(U)$ with the
following property: $\forall u \in U : X(u) = x$, it must be that
$Y(u) = f_\gamma(\Gamma(u))$, where $f_\gamma : \Gamma(U) \rightarrow
{\mathbb{B}}$ is the probe of $\Gamma$'s range that equals $1$ iff
$\Gamma(u) = \gamma$.

Intuitively, to have $C > \Gamma$ means that if the image value of
$\Gamma$ is expressed as a list of answers to questions of the form
``Does $\Gamma(u) = \gamma$?'', then we can set up the device so that
it will guaranteedly correctly conclude any particular answer in that
list. Alternatively, the requirement that there be an appropriate $x$
for any probe function of $\Gamma$ can be viewed as shorthand; in the
definition of inference we are considering the ability of a device to
correctly answer any member of a list of binary-valued questions, a
set that is ``generated'' by $\Gamma$. So weak-inference is a
worst-case definition: if a device $C$ weakly infers $\Gamma$, then no
matter what probe $f \in \pi(\Gamma)$ a malicious demon might choose,
the scientist could guarantee that $Y = f(\Gamma)$ by choosing an
associated value $x$ for the value of $X$.

To illustrate this, consider again Ex. 1. Identify the $Y$ in Def. 3
with the $Y$ in Ex. 1, and similarly identify the $\Gamma$'s with
each other. Then identify the function $X$ in Def. 3 as the 
product of functions, $\chi \otimes Q$. $(X, Y)$ specifies a device
$C$. The functions $f_K$ in Ex. 1 are the probes in $\pi(\Gamma)$. So
if $C > \Gamma$, then the aggregate system of scientist and
observation apparatus can observe $S(t_2)$. Note that $\zeta$ ends up
being irrelevant.  In essence, it serves as a conduit to transfer
information into the scientist's brain.

In the many-worlds definition of an observation, any particular result
of the observation is identified with a solitary worldline
$u$. Intuitively, this might be worrisome; a solitary $u$ is just a
single point in a space, with no intrinsic mathematical structure. The
properties of such a single point can be drastically modified by an
appropriate isomorphism over $U$. In particular, as has been pointed
out by many authors, in the many-worlds definition what gets
``observed'' can be modified if one changes the basis of $U$. (This is
one of the major motivations for the work on
decoherence~\cite{zure03,zeh70}.)

However if a scientist makes an observation, then that scientist
{\it{could}} provide the value of any (binary-valued) function of the
result of the observation, if they were asked to. So formally
requiring that the scientist be able to provide such values doesn't
preclude real-world instances of observation.  At the same time,
adding such a requirement has substantial consequences. In fact, it
drives many of the results presented below concerning weak
inference. This is why this requirement is incorporated into the
definition of weak inference.  In other words, it is why the
definition of weak inference inherently involves multiple worldlines
$u$, in contrast to the many-worlds definition of observation. 

See Sec.~\ref{sec:philo} for a discussion of the philosophical aspects
of weak inference.  The relation between weak inference and the theory
of knowledge functions~\cite{auma99,aubr95,bibr88,futi91} is briefly
discussed in Sec.~\ref{sec:sad}.  App. B contains a discussion of how
unrestrictive the definition of weak inference is. Finally, some
alternative definitions of devices and weak inference are considered
in App. C.

\subsection{Elementary results concerning weak inference}

We say that a device $C_1$ infers a set of functions if it infers
every function in that set. We also say $C_1$ infers a device $C_2$
iff $C_1 > Y_2$. In general inference among devices is
non-transitive. In addition we have the following elementary
properties of devices:

$ $

\noindent {\bf{Proposition 1:}} Let $\{\Gamma_i\}$ be a set of
functions with domain $U$ and $W \subset U$.

{\bf{i)}} If $\forall i$, $|\Gamma_i(W)| \ge 2$, then there is a
device over $U$ that infers $\{\Gamma_i\}$.

{\bf{ii)}} For any device $C$, there is a binary-valued function that
$C$ does not infer.

$ $

\noindent Prop. 1(ii) means in particular that there are sets $\{\Gamma_i\}$  such
that no device can infer every function in that set.  

In a limited sense, when applied to prediction (cf. Ex. 1),
Prop. 1(ii) means that Laplace was wrong: even if the universe were a
giant clock, he would not have been able to reliably predict the
universe's future state before it occurred.{\footnote{Similar
conclusions have been reached previously~\cite{mack60,popp88}. However
in addition to being limited to the inference process of prediction,
that earlier work is quite informal. Furthermore, it unknowingly
disputes well-established results in engineering. For example, the
claim in~\cite{mack60} that ``a prediction concerning the narrator's
future ... cannot ... account for the effect of the narrator's
learning that prediction'' is refuted by adaptive control theory and
Bellman's equations. Similarly, those with training in computer
science will recognize statements (A3), (A4), and the notion of
``structurally identical predictors'' in~\cite{popp88} as formally
meaningless.}} Viewed differently, Prop. 1(ii) means that regardless
of noise levels and the dimensions and other characteristics of the
underlying attractors of the physical dynamics of various systems,
there cannot be a time-series prediction algorithm~\cite{lama94} that
is always correct in its prediction of the future state of such
systems.

Note that time does not appear in Def. 3's model of a prediction
system.  So in particular in Ex. 3 we could have $t_3 < t_2$ --- so that
the time when the computer provides its prediction is $after$ the
event it is predicting --- and the impossibility result of Prop. 1(ii)
still holds (cf. Ex. 4). Moreover, the program that is input to the
prediction computer via the value of $\chi$ could even contain the
value that we want to predict. Prop. 1(ii) would still mean that the
conclusion that the computer's user comes to after reading the
computer's output cannot be guaranteed to be correct.

This is all true even if the computer has super-Turing capability, and
does not derive from chaotic dynamics, physical limitations like the
speed of light, or quantum mechanical limitations. Indeed, when
applied to an observation apparatus like in Ex. 1, Prop. 1(ii) can be
viewed as a sort of non-quantum mechanical ``uncertainty principle'',
establishing that there is no general-purpose, infallible observation
device. (See also Prop. 6 below, which is perhaps more closely
analogous to the uncertainty principle.) In addition, when applied to
the recording apparatus of Ex. 4, Prop. 1(ii) means that there is no
general-purpose, infallible recording device.

To illustrate this in more detail, consider the relatively simple
scenario where $C$ is a computer making a prediction at time $t$ about
the state of the (deterministic, classical) universe at $t' > t$. Let
$G$ be the set of all time-$t$ states of the universe in which $C$'s
output display is $+1$. The laws of physics can be used to evolve $G$
forward to time $t'$. Label that evolved set of time-$t'$ states of
the universe as $H$. Let $\Gamma$ be the binary-valued question,
``does the state of the universe at $t'$ lies outside of $H$?''.

There is no information concerning $H$ that can be programmed into $C$
at some time $t^- < t$ that guarantees that the resultant prediction
that $C$ makes at $t$ is a correct answer to that question.  This is
true no matter what $t^-$ is, i.e., no matter how much time $C$ has to
run that program before making its answer at time $t$. It is also true
no matter how much time there is between $t'$ and $t$. It is even true
if the program with which $C$ is initialized explicitly gives the
correct answer to the question.

Similar results hold if $t' < t$. In particular, such results hold if
$C$ is an observation device that we wish to configure so that at time
$t$ it correctly completes an observation process saying whether the
universe was outside of $H$ at time $t'$. We can even have $t'$ be
earlier than the time when $C$ is set up. In this case, $C$ is a
recording system that contains information about the past and we wish
to query it about whether the universe was outside of $H$ at
$t'$. See~\cite{wolp01} for further discussion of these points.

While these limitations are unavoidable, often they are not relevant,
in that we are not interested in whether a device infers an arbitrary
set of functions.  Instead, often we are interested in whether a
devices infers some specified subset of all functions.  Prop. 1(i)
addresses that situation. In particular, given our assumption that any
function over $U$ must contain at least two values in its range, it
immediately implies the following:

$ $

\noindent {\bf{Corollary 1:}} 

{\bf{i)}} Let $\{\Gamma_i\}$ be a set of functions with domain $U$ and
$W \subset U$.  If $\forall i$, $\Gamma_i(U) = \Gamma_i(W)$, 

$\;\;\;$ then there is a device that infers $\{\Gamma_i\}$.

{\bf{ii)}} For any function $\Gamma$ with domain $U$ there is a device
that infers $\Gamma$.

$ $

\noindent Another implication of Prop. 1(i) is the following:

$ $

\noindent {\bf{Corollary 2:}} Let $C = (X, Y)$ be a device over
$U$ where the partition induced by $X$ is a fine-graining of the
partition induced by $Y$.  Then $|X(U)| > 2$ iff there is a function
that $C$ infers.

$ $

Prop. 1(ii) tells us that any inference device $C$ can be ``thwarted''
by an associated function. However it does not forbid the possibility
of some second device that can infer that function that thwarts $C$.
To analyze issues of this sort, and more generally to analyze the
inference relationships within sets of multiple functions and multiple
devices, we start with the following definition:

$ $

\noindent{\bf{Definition 4:}} Two devices $(X_1, Y_1)$ and 
$(X_2, Y_2)$ are {\bf{(setup) distinguishable}} iff $\forall \; x_1,
x_2, \; \exists \; u \in U$ s.t. $X_1(u) = x_1, X_2(u) = x_2$.

$ $

No device is distinguishable from itself. Distinguishability
is non-transitive in general. Having two devices be
distinguishable means that no matter how the first device is set up,
it is always possible to set up the second one in an arbitrary
fashion; the setting up of the first device does not preclude any
options for setting up the second one. Intuitively, if two devices are not distinguishable, then the setup function of one of the devices is partially ``controlled" by the setup function of the other one. In such a situation, they are not two fully separate, independent devices.

By choosing the negation probe $f(y \in {\mathbb{B}}) = -y$ we see
that no device can weakly infer itself. We also have the following:

$ $

\noindent {\bf{Theorem 1:}} No two distinguishable devices can weakly
infer each other.

$ $

\noindent Thm. 1 says that no matter how clever we are in designing a pair of inference
devices, so long as they are distinguishable from each another, one of
them must thwart the other, providing a function that the other
device cannot infer. Whereas the impossibility result of Prop. 1(ii)
relies on constructing a special function $\Gamma$ matched to $C$, the
implications of Thm. 1 are broader, in that they establish that a
whole class of functions cannot be inferred by $C$ (namely the
conclusion functions of devices that are distinguishable from $C$ and
also can infer $C$). It is important to note that the
distinguishability condition is crucial to Thm. 1; mutual weak
inference can occur between non-distinguishable devices.

$ $

\noindent {\bf{Example 5:}} Consider a rectangular
grid of particle pairs, each pair consisting of a yellow particle and
a purple particle. Say that all particles can either be spin up or
spin down. Write the spin of the purple particle at grid location
$(i,j)$ as $s^p(i, j)$, and the spin of the yellow particle there as
$s^y(i,j)$.

Such a grid is a set $U$ consisting of all quadruples $\{i, j,s^p(i,
j), s^y(i, j)\}$. Assume there are at least two $i$ values, and at
least one purple spin is up and at least one is down. Then we can
define a ``purple inference device'' $C^p$ by $X^p(i, j, s^p(i, j),
s^y(i, j)) \triangleq i$ and $Y^p(i, j, s^p(i, j), s^y(i, j))
\triangleq s^p(i, j)$. Similarly, a ``yellow inference device'' can be
defined by $X^y(i, j, s^p(i, j), s^y(i,j)) \triangleq j$ and $Y^y(i,
j, s^p(i, j), s^y(i, j)) \triangleq s^y(i,j)$ (assuming there are at
least two $j$'s and at least one yellow particle is spin up and at
least one is spin down).

These two devices are distinguishable. In addition, $C^p > C^y$ if
there is some $i'$ such that $s^p(i', j) = s^y(i', j)$ for all $j$,
and also some $i''$ such that $s^p(i'', j) = -s^y(i'', j)$ for all
$j$. In such a situation we can set up the purple device with a value
($i'$) that guarantees that its conclusion correctly answers the
question, ``Does $s^y$ point up?''. Similarly, we can set it up with a
value that guarantees that its conclusion correctly answers the
question, ``Does $s^y$ point down?''. 

However if there is such an $i'$ and $i''$, then clearly there cannot
also be both a value $j'$ and a value $j''$ that the yellow inference
device can use to answer whether $s^p$ points up and whether $s^p$
points down, respectively. This impossibility holds regardless of the
size of the grid and the particular pattern of yellow and purple
particles on the grid. Thm. 1 generalizes this impossibility result.

$ $

As a general comment, the definition of what it means for a device to
infer $\Gamma$ can be re-expressed in terms of the pre-images in $U$
of $\Gamma$, $\{\Gamma^{-1}(\gamma) :
\gamma \in \Gamma(U)\}$.{\footnote{Writing it out, if $C$ infers $\Gamma$, then
for all $\forall \; \gamma \in \Gamma(U), \exists \; x \in X(U)$ such
that $[X^{-1}(x) \cap Y^{-1}(1)] = [X^{-1}(x)
\cap \Gamma^{-1}(\gamma)]$.}} Now in this paper we only consider weak
inference of $\Gamma$'s that are functions. So none of those
pre-images of $\Gamma$ intersect the others; they comprise a partition
of $U$. However more generally, one might be interested in inference
of $\Gamma$ when some of the pre-images of $\Gamma$ have non-empty
intersection with one another. For example, one might wish to observe
if some physical variable is in the range $[0, 10]$, the range $[5,
20]$, or the range $[15, 30]$.  Formally, the generalization to
overlapping pre-images of $\Gamma$ arises by allowing $\Gamma$ to be a
correspondence rather than a function. The generalization of the
formalism to explicitly accommodate such correspondences is beyond the
scope of this paper. Note though that since devices are pairs of
functions, that generalization is not relevant for much of the
analysis concerning the inference of one device by another.

\section{Turing machines, Universal Turing machines, and inference}

There are several connections between inference and results in
computer science~\cite{houl79}. In this section we introduce some
elementary concepts for exploring those connections.

\subsection{Turing machines and inference}

Consider a deterministic Turing Machine (TM) and write its internal
state at iteration $t$ as $g(t)$, with the state of its tape then
being written as $h(t)$. So the operation of the TM on a particular
initial value of its tape $h(t_0)$ produces an infinite sequence
\{$h(t_0), g(t_0), h(t_0 + 1), g(t_0 + 1), \ldots$\}. (If $g(t)$ is
the halt state, then for completeness we define $g(t') = g(t), h(t') =
h(t) \; \forall t' > t$.) Which such sequence the TM executes is
determined by the value $h(t_0)$ (assuming a default value for
$g(t_0)$).

Next take $U$ to be the set of worldlines consistent with the laws of
physics in our universe (and no other worldlines). Hypothesize that it
is consistent with those laws of physics to have some particular TM
$T$ be physically instantiated in our universe, with iteration number
$t$ corresponding to time in some particular reference frame. Then
which sequence $T$ actually executes can be cast as a projection
function of the worldline $u \in U$. (Recall that worldlines extend
across all time.)  Accordingly we can identify any $T$ as a function
$\Gamma$ with domain $U$. The set of all possible sequences of $T$
that can occur in our universe is simply a set of functions $\Gamma$.

To be more precise, fix $t_0$, and let $H^T$ be the set of all
possible initial (time $t_0$) values of $T$'s tape. Define $M^T$ as
the map by which $T$ takes $h(t_0) \in H^T$ to the associated infinite
sequence
\{$h(t_0), g(t_0), h(t_0 + 1), g(t_0 + 1), \ldots$\}. $M^T$ can be viewed as defining
$T$. Equivalently, we can express $T$ as a function over $U$,
$\Gamma^T$: $\Gamma^T$ projects every $u \in U$ in which $T$ has
initial tape state $h \in H^T$ to $M^T(h)$. $M^T$ and $\Gamma^T$ have
the same range (namely the set of all sequences that $T$ can
generate), but different domains ($H^T$ and $U$, respectively).

Now construct an inference device $C^T \equiv (X^T, Y^T)$ where
$X^T(U) \equiv \{(h, f) : h \in H^T, f \in \pi(\Gamma^T)\}$. Write the
two components of any value $X^T(u)$ as $X^T_h(u)$ and $X^T_f(u)$,
where $X^T_h(u)$ is defined to be the value $h(t_0)$ for the TM $T$
when the worldline is $u$. So $X^T_h$ ``initializes'' the TM. Note that
the second component of $X$, $X^T_f$, maps $u$ onto a space of
functions over $U$ (namely, the space $\pi(\Gamma)$).  Finally, define
$Y^T : u \rightarrow 1$ iff $X^T_f(u)[M^T(X^T_h(u))] = 1$.

If $X^T$ is set up to be a particular initial state of $T$'s tape,
together with a particular probe concerning the resultant sequence of
internal and tape states, then for any $u$ the conclusion $Y^T(u)$ is
the actual value of that probe for the sequence of internal and tape
states specified in $u$. Since probes are simply a way to imbue the
conclusion of the device with semantic meaning (recall Ex. 3 in
Sec.~\ref{sec:examples}), this means we can view $C$ as equivalent to
$T$. In particular, $C^T$ infers the TM, i.e., $C^T >
\Gamma^T$.

We can generalize this example, to identify inference devices in
general as analogs of TM's, with inference being the analog of
TM-style computation. All of the impossibility results presented above
apply to these analogs of TM's. To illustrate this, Prop. 1(ii) can be
taken to mean that for any such inference-based analog of a TM, there
is some function that the device cannot ``compute''. In particular,
this is true for the device $C^T$ that essentially equals the TM
$T$. In this, Prop. 1(ii) can be viewed as the analog for inference
devices of the Halting theorem, which concerns TM's.  Moreover, this
reasoning concerning physical realizations of TM's applies just as
well to other members of the Chomsky hierarchy besides TM's, providing
us with ``halting theorems'' for those other members.

As a final comment on the relation between inference and TM-style
computation, note that inference by a device $C$ is not a form of
counter-factual ``computation''. Inference by $C$ does not compute the
answer to a question of the form ``If \{axioms\} then
\{implications\}'', unless there is some $x$ such that ``\{axioms\}'' actually holds for all $u
\in U$ that $C$ induces by setting $X(u) = x$. In
particular, if in our universe there is no physical instantiation of
some particular TM, then there is no device in our universe whose
inference is computationally equivalent to that TM.

\subsection{Universal Turing machines and inference}

Now we investigate how to define an analog of Universal Turing
Machines (UTM's) for inference devices. More precisely, we consider
how to define what it means for one device $C_1$ to emulate the
inference process of another device $C_2$. (Just like a UTM emulates
the computational process of another TM.)  One natural desideratum for
such a definition is that for $C_1$ to ``emulate'' $C_2$ implies, at a
minimum, that $C_1 > C_2$. So for example, if the two devices are both
being used for prediction, this would mean that $C_1$ can correctly
predict what prediction $C_2$ will make (whether or not that
prediction by $C_2$ is itself correct).

However we want $C_1$ able to do more than infer the value of
$Y_2(u)$; we want $C_1$ able to emulate the entire mapping taking any
$x_2$ to the associated value(s) $Y_2(X_2^{-1}(x_2))$. We want $C_1$
able to infer what inference $C_2$ might make for $any$ setup value
$x_2$, not just the inference that $C_2$ makes for the members of a
set $X_2[X^{-1}_1(x_1)]$ picked out by some particular $x_1$.  This
means that all $x_2$'s must be allowed.

One way to formalize this second desideratum is to require that $C_1$
can infer $C_2$ using a setup value that forces a unique $x_2$, and
can do so for any desired $x_2$. More precisely, consider a particular
case where we want $C_1$ to emulate the inference performed by $C_2$
when $X_2(u) = x_2$. We can do this if $C_1$ infers $Y_2$, while the
value $x_1$ used in that inference guarantees that $X_2(u) = x_2$.
That guarantee means that $C_1$ infers the conclusion of $C_2$ when
$C_2$ has the setup value $x_2$. Given this interpretation of what it
means for $C_1$ to emulate $C_2$ when $X_2(u) = x_2$, to have $C_1$
emulate $C_2$ in full simply means that we require that such emulation
be possible for any $x_2 \in X_2(U)$. So formally, we require that
$\forall f
\in \pi(Y_2),
\forall x_2, \exists x_1$ such that $X_1 = x_1 \Rightarrow X_2 = x_2, Y_1 =
f(Y_2)$.

A second formalization takes the opposite approach, and stipulates
that the value $x_1$ used by $C_1$ to infer $C_2$ places no
restrictions on $x_2$ whatsoever.  Formally, this means that $\forall
f \in \pi(Y_2),
\forall x_2, \exists x_1$ such that $X_1^{-1}(x_1) \cap X_2^{-1}(x_2)
\ne \varnothing$ and $X_1 = x_1 \Rightarrow Y_1 =
f(Y_2)$.

In analogy with UTM's, one might say that under the first
formalization $C_1$ specifies the ``input tape'' to $C_2$ for which
$C_1$ will emulate $C_2$, and then successfully carries out that
emulation, i.e., successfully ``computes'' what $C_2$ will produce in
response to that input tape. To do this though $C_1$ must interfere
with $C_2$, forcing it to have that desired input tape. In contrast,
under the second formalization, there is no requirement that $X_1$
force a particular value of $X_2$. In particular, the second
formalization is obeyed if $\forall f \in \pi(Y_2), \; \exists x_1$
such that $X_1 = x_1 \Rightarrow Y_1 = f(Y_2)$ while at the same time
$X_1^{-1}(x_1) \cap X_2^{-1}(x_2) \ne \varnothing \; \forall x_2$. In
such a situation, $C_1$ can emulate $C_2$ using an $x_1$ that doesn't
reflect how $C_2$ is set up. (Physically, this usually requires that
the system underlying $C_1$ must be coupled with the system underlying
$C_2$ at some time, so that $x_2$ can be made known to $C_1$.)

Despite this apparent difference, these two formalizations of our
second desideratum reflect the same underlying mathematical
structure. To see this, define a composite device $C' = (X', Y')$
where $X' : u \rightarrow (X_1(u), X_2(u))$ and $Y' = Y_1$. Then under
our second formalization of ``emulation'', for $C_1$ to emulate $C_2$
implies that $\forall f \in \pi(Y_2), \forall x_2, \exists x'$ such
that $X'^{-1}(x') \cap X^{-1}_2(x_2) \ne \varnothing$ and $X' = x'
\Rightarrow X_2 = x_2, Y' = f(Y_2)$.  However $X'^{-1}(x') \cap
X^{-1}_2(x_2) \ne \varnothing$ means that $X' = x' \Rightarrow X_2 =
x_2$, by definition of $X'$.  So this second formalization of what it
means for $C_1$ to emulate $C_2$ stipulates a relation between $C'$
and $C_2$ that is identical to the relation between $C_1$ and $C_2$
under the first formalization. In this sense, our second formalization
reduces to our first. Accordingly, we concentrate on the first
formalization, and make the following definition:

$ $

\noindent {\bf{Definition 5:}} A device $(X_1, Y_1)$ {\bf{strongly
infers}} a device $(X_2, Y_2)$ iff $\forall \; f \in \pi(Y_2)$ and all
$x_2$, $\exists \; x_1$ such that $X_1 = x_1 \Rightarrow X_2 = x_2,
Y_1 = f(Y_2)$.

$ $

\noindent If $(X_1, Y_1)$ strongly infers $(X_2, Y_2)$ we write $(X_1,
Y_1) \gg (X_2, Y_2)$.{\footnote{Note that there are only two probes of
$Y_2$, the identity probe $f(y_2) = y_2$ and the negation probe,
$f(y_2) = -y_2$. Indicate those two probes by $f = 1$ and $f = -1$,
respectively. Then we can express $X_1 = x_1 \Rightarrow X_2 = x_2,
Y_1 = f(Y_2)$ in set-theoretic terms, as $X_1^{-1}(x_1) \subseteq
X_2^{-1}(x_2) \; \cap \; (Y_1Y_2)^{-1}(f)$, where $Y_1Y_2$ is the
function $u \in U \rightarrow Y_1(u) Y_2(u)$.}\label{ft:2}} See App. B
for a discussion of how minimal the definition of strong inference
really is.

Say we have a TM $T_1$ that can emulate another TM $T_2$, e.g., $T_1$
is a UTM. This means that $T_1$ can calculate anything that $T_2$
can. The analogous property holds for strong and weak inference. In
addition, like UTM-style emulation (but unlike weak inference), strong
inference is transitive. These results are formalized as follows:

$ $

\noindent {\bf{Theorem 2:}} Let $C_1$, $C_2$  and $C_3$ be a set of
inference devices over $U$ and $\Gamma$ a function over $U$. Then:

{\bf{i)}} $C_1 \gg C_2$ and $C_2 > \Gamma$ $\Rightarrow$ $C_1 >
\Gamma$.

{\bf{ii)}} $C_1 \gg C_2$ and $C_2 \gg C_3$ $\Rightarrow$ $C_1 \gg C_3$.

$ $

Strong inference implies weak inference, i.e., $C_1 \gg C_2
\Rightarrow C_1 > C_2$.  We also have the following strong inference
analogs of Prop. 1(ii) and Coroll. 1 (which concerns weak inference):

$ $

\noindent {\bf{Proposition 2:}} Let $C_1$ be a device over $U$.

{\bf{i)}} There is a device $C_2$ such that $C_1 \not \gg C_2$.

{\bf{ii)}} Say that $\forall \; x_1$, $|X_1^{-1}(x_1)| >
2$.  Then there is a device $C_2$ such that $C_2 \gg C_1$.

$ $

Recall that the Halting problem concerns whether there is a UTM $T$
with the following property: Given any TM $T'$ and associated input
string $s'$, if $T'$ and $s'$ are encoded as an input string to $T$,
then $T$ always correctly decides whether $T'$ halts on input
$s'$. The Halting theorem then says that there can be no such UTM
$T$. Intuitively, Prop. 2(i) can be viewed as an analog of this
theorem, in the context of inference. (See also Prop. 7 below.)

In general we are not interested in whether a device can strongly
infer an arbitrary set of other devices, but rather with the strong
inference relationships among the members of a particular set of
devices.  Just like with weak inference, no device can strongly infer
itself. This can be generalized to concern a set of multiple devices
as follows:

$ $

\noindent {\bf{Theorem 3:}} No two devices can strongly
infer each other.

$ $

\noindent Note that Thm. 3 does not require distinguishability, in
contrast to Thm. 1.

\section{Inference Complexity}
\label{sec:inf_compl}

In computer science, given a TM $T$, the Kolmogorov complexity of an
output string $s$ is defined as the length of the smallest input
string $s'$ that when input to $T$ produces $s$ as output. To
construct our inference device analog of this, we need to define the
``length'' of an input region of an inference device $C$. To do this,
we assume we are given a measure $d\mu$ over $U$, and for simplicity
restrict attention to functions $G$ over $U$ with countable
range. Then we define the {\bf{length}} of $g \in G(U)$ as -ln$[\int
d\mu \; G^{-1}(g)]$, i.e., the negative logarithm of the volume of all
$u \in U$ such that $G(u) = g$. We write this length as
${\mathscr{L}}_C(g)$, or just ${\mathscr{L}}(g)$ for
short.{\footnote{If $\int d\mu \; 1 = \infty$, then we instead work
with differences in logarithms of volumes, evaluated under an
appropriate limit of $d\mu$ that takes $\int d\mu \; 1 \rightarrow
\infty$. For example, we might work with such differences when $U$ is
taken to be a box whose size goes to infinity. This is just the usual
physics trick for dealing with infinite volumes. \label{foot:vol}}}

$ $

\noindent {\bf{Definition 6:}} Let $C$ be a device and $\Gamma$ a function over
$U$ where $X(U)$ and $\Gamma(U)$ are countable and $C > \Gamma$. The
{\bf{inference complexity}} of $\Gamma$ with respect to $C$ is defined
as
\begin{eqnarray*}
{\mathscr{C}}(\Gamma \mid C) \;\;&\triangleq& \;\; \sum_{f \in \pi(\Gamma)}
	{\mbox{min}}_{x : X = x \Rightarrow Y = f(\Gamma)}
			 [{\mathscr{L}} (x)].
\end{eqnarray*}

$ $

\noindent The inference complexity of $\Gamma$ with respect to $C$ is the sum of
a set of ``complexities'', one for each probe of $\Gamma$,
$f$. Loosely speaking, each of those complexities is the minimal
amount of Shannon information that must be imposed in $C$'s setup
function in order to ensure that $C$ correctly concludes what value
$f$ has. In particular, if $\Gamma$ corresponds to a potential future
state of some system $S$ external to $C$, then ${\mathscr{C}}(\Gamma \mid
C)$ is a measure of how difficult it is for $C$ to predict that future
state of $S$. Loosely speaking, the more sensitively that future state
depends on current conditions, the more complex is the computation of
that future state.

$ $

\noindent {\bf{Example 6:}} Consider a
conventional real-world computer, with a subsection of its RAM set
aside to contain the program it will run, and a separate subsection
set aside to contain the conclusion that the program will produce. Say
the total number of bits in the program subsection of the RAM is $2^k
+ k$ for some integer $k$. Refer to any set of $2^k + k$ bits as a
``complete string''; the set of all complete strings is the set of all
possible bit strings in the program subsection of the RAM.

Let $\Sigma^k$ be the set of all bit strings $s$ consisting of at
least $k$ bits such that the first $k$ bits are a binary encoding of
the total number of bits in $s$ beyond those first $k$ bits. So every
element of $\Sigma^k$ can be read into the beginning of the RAM's
program subsection.  For any $s \in \Sigma^k$ define an associated
``partial string'' as the set of all complete strings whose first bits
are $s$. Intuitively, for any such complete string, all of its bits
beyond $s$ are ``wild cards''. (Such partial strings are just the
``files'' of real-world operating systems.) With some abuse of
terminology, when we write ``$s$'' we will sometimes actually mean the
partial string that $s$ specifies.

We can identify a particular program input to the computer as such a
partial string in its program subsection.  If we append certain bits
to such an $s$ (modifying the contents of the first $k$ bits
appropriately) to get a new longer program partial string, $s'$, the
set of complete strings consistent with $s'$ is a proper subset of the
set of complete strings consistent with $s$.

Define the length of a partial string $s$ as the negative of the
logarithm of the number of complete strings that have $s$ at their
beginning, minus $k$.  This matches the usual definition of the length
of a string used in computer science.  In particular, if $s'$ contains
$n$ more bits than does $s$, then there are $2n$ times as many
complete strings consistent with $s$ as there are consistent with
$s'$. Accordingly, if we take logarithms to have base 2, the length of
$s'$ equals the length of $s$, plus $n$.

Now view our physical computer as an inference device, with $U$ the
Cartesian product of the set of all possible bit strings in the RAM of
the computer together with some countable-valued variables concerning
the world outside of the computer. Refer to the components of any $u
\in U$ specifying the bit string in the program subsection of the RAM
as the ``program subsection of $u$'', and similarly for the ``conclusion
subsection of $u$''.

For the computer to be an inference device means that the conclusion
subsection of $u$ consists of a single bit, i.e., $Y$ maps all $u \in
U$ to the (bit) value of the conclusion subsection of the computer's
RAM as specified by $u$. For all $u \in U$, have $X(u)$ be the bit
string at the beginning of the program subsection of $u$ whose length
is given by the first $k$ bits of that program subsection of $u$. So
$x$ is a partial string of the RAM's program subsection. In general,
there are many sets each consisting of multiple $u \in U$ that have
the same image under $X$, i.e., there are many $x$ such that
$X^{-1}(x)$ consists of multiple elements. If we adopt the uniform
point measure $d\mu$, then ${\mathscr{L}}(x)$ is just the negative
logarithm of the number of such elements in $X^{-1}(x)$, i.e., the
length of the partial string $x$ in the program subsection of the
computer's RAM.

Now say we want our computer to make a prediction concerning the value
of $\Gamma(U)$, one of the variables associated with the world outside
of the computer. As usual, we interpret this to mean that for any
$\gamma \in \Gamma(U)$, there is some partial string we can read into
the computer's program subsection that contains enough information
concerning $\Gamma$ and the state of the world so that the computer's
conclusion will correctly say whether $\Gamma(u) = \gamma$.  The
inference complexity of that prediction of $\Gamma$ is the sum, over
all such probes $f$ of $\Gamma$, of the length of the shortest partial
string in the computer's program subsection that cause it to correctly
conclude the value of $f$.

$ $

The min over $x$'s in Def. 6 is a direct analog of the min in the
definition of Kolmogorov complexity (there the min is over those
strings that when input to a particular UTM result in the desired
output string). A natural modification to Def. 6 is to remove the
min by considering all $x$'s that cause $Y = f(\Gamma)$, not just of
one of them:
\begin{eqnarray*}
{\hat{{\mathscr{C}}}}(\Gamma \mid C) \;\;&\triangleq& \;\; \sum_{f \in
\pi(\Gamma)} -{\mbox{ln}} \left[\; \mu \left(\cup_{x :  X
= x \Rightarrow Y = f(\Gamma)} X^{-1}(x) \right) \;\right]  \nonumber \\
&=& \sum_{f \in \pi(\Gamma)} -{\mbox{ln}} \left[\sum_{x : X = x \Rightarrow
Y = f(\Gamma)} e^{-{\mathscr{L}}(x)}\right],
\end{eqnarray*}
where the equality follows from the fact that for any $x, x' \ne x$,
$X^{-1}(x) \cap X^{-1}(x') = \varnothing$.  The argument of the ln in
this modified version of inference complexity has a direct analog in
TM theory: The sum, over all input strings $s$ to a UTM that generate
a desired output string $s'$, of $2^{-n(s)}$, where $n(s)$ is the bit
length of $s$.

We now bound how much more complex a function can appear to $C_1$ than
to $C_2$ if $C_1$ can strongly infer $C_2$.

$ $

\noindent {\bf{Theorem 4:}} Let $C_1$ and $C_2$ be two  devices and
$\Gamma$ a function over $U$ where $\Gamma(U)$ is finite, $C_1 \gg
C_2$, and $C_2 > \Gamma$. Then
\begin{eqnarray*}
{\mathscr{C}}(\Gamma \mid C_1) - {\mathscr{C}}(\Gamma \mid C_2) \;\;\; &\le& \;\;\;
|\Gamma(U)| \; {\mbox{max}}_{x_2} {\mbox{min}}_{x_1 : X_1 = x_1
\Rightarrow X_2 = x_2, Y_1 = Y_2} [{\mathscr{L}}(x_1) -
{\mathscr{L}}(x_2)] .
\end{eqnarray*}
\noindent 

$ $

\noindent Note that since ${\mathscr{L}}(x_1) - {\mathscr{L}}(x_2) =
{\mbox{ln}}[\frac{X_2^{-1}(x_2)} {X_1^{-1}(x_1)}]$, the bound in
Thm. 4 is independent of the units with which one measures volume in
$U$. (Cf. footnote ~\ref{foot:vol}.) Furthermore, recall that $X_1 =
x_1 \Rightarrow X_2 = x_2, Y_1 = Y_2$ iff $X_1^{-1}(x_1) \subseteq
X_2^{-1}(x_2) \; \cap \; (Y_1Y_2)^{-1}(1)$. (Cf. footnote~\ref{ft:2}.) 
Accordingly, for all $(x_1, x_2)$ pairs arising in the bound in
Thm. 4, $\frac{X_2^{-1}(x_2)} {X_1^{-1}(x_1)} \ge 1$. So the bound in
Thm. 4 is always non-negative.

An important result in the theory of UTM's is an upper bound on the
difference between the Kolmogorov complexity of a string using a
particular UTM $T_1$ and its complexity if using a different UTM,
$T_2$. This bound is independent of the computation to be performed,
and can be viewed as the Kolmogorov complexity of $T_1$ emulating
$T_2$.

The bound in Thm. 4 is the analog of this UTM result, for inference
devices. In particular, the bound in Thm. 4 is independent of all
aspects of $\Gamma$ except the cardinality of $\Gamma(U)$.
Intuitively, the bound is $|\Gamma(U)|$ times the worst-case amount of
``computational work'' that $C_1$ has to do to ``emulate'' $C_2$'s
behavior for some particular value of $x_2$.

\section{Realities and copies of devices}

In this section the discussion is broadened to allow sets of many
functions to be inferred and / or inference devices. Some of the
philosophical implications of the ensuing results are then discussed.

\subsection{Formal results}

To analyze relationships among multiple devices and functions, define
a {\bf{reality}} as a pair $(U; \{F_\phi\})$ where $U$ is a space and
$\{F_\phi\}$ is a (perhaps uncountable) non-empty set of functions all
having domain $U$. We will sometimes say that $U$ is the {\bf{domain}}
of the reality. We are particularly interested in {\bf{device
realities}} in which some of the functions are binary-valued, and we
wish to pair each of those functions uniquely with some of the other
functions. Such realities can be written as the triple $(U;
\{(X_\alpha, Y_\alpha)\}; \{\Gamma_\beta\}) \equiv (U; \{C_\alpha\};
\{\Gamma_\beta\})$ where $\{C_\alpha\}$ is a set of devices over $U$
and $\{\Gamma_\beta\}$ a set of functions over $U$.

Define a {\bf{universal device}} as any device in a reality that can
strongly infer all other devices and weakly infer all functions in
that reality. Thm. 3 means that no reality can contain more than one
universal device. So in particular, if a reality contains at least one
universal device, then it has a unique natural choice for an inference
complexity measure, namely the inference complexity with respect to
its (unique) universal device. (This contrasts with Kolmogorov
complexity, which depends on the arbitrary choice of what UTM to use.)

It is useful to define the {\bf{reduced form}} of a reality $(U;
\{F_\phi\})$ as the range of $\bigotimes_\phi F_\phi$. Expanding, this
equals $\cup_{u \in U} [\varprod_\phi F_\phi](u)$, the union over all
$u$ of the tuples formed by a Cartesian product, running over all
$\phi$, of the values $F_\phi(u)$. In particular, the reduced form of a
device reality is the set of all tuples $([x_1, y_1], [x_2, y_2],
\ldots; \gamma_1, \gamma_2, \ldots)$ for which $\exists \; u \in U$
such that simultaneously $X_1(u) = x_1, Y_1(u) = y_1, X_2(u) = x_2,
Y_2(u) = y_2, \ldots ; \Gamma_1(u) =
\gamma_1, \Gamma_2(u) = \gamma_2, \ldots$.

As an example, take $U$ to be the set of all worldlines consistent
with the laws of physics (and no other worldlines). So for example, if
one wants to consider a universe in which the laws of physics are
time-reversible and deterministic, then we require that no two
distinct members of $U$ can intersect. Similarly, properties like
time-translation invariance can be imposed on $U$, as can more
elaborate laws involving physical constants. Which such particular
properties of $U$ are imposed depends on what the laws of physics are.

Next, have \{$\Gamma_\beta$\} be a set of physical characteristics of
the universe, each characteristic perhaps defined in terms of the
values of one or more physical variables at multiple locations and/or
multiple times. Finally, have \{$C_\alpha$\} be all prediction /
observation systems concerning the universe that all scientists might
ever be involved in.

This example is the conventional way to interpret our universe as a
reality. In this example the laws of physics are embodied in $U$. The
implications of those laws for the relationships among the scientist
devices \{$C_\alpha$\} and the other characteristics of the universe
\{$\Gamma_\beta$\} is embodied in the reduced form of the
reality. Viewing the universe this way, it is the $u \in U$,
specifying the universe's state for all time, that has ``physical
meaning''. The reduced form instead is a logical implication of the
laws of the universe. In particular, our universe's $u$ picks out the
tuple $[\varprod_\alpha C_\alpha (u)] \times [\varprod_\beta
\Gamma_\beta(u)]$ from the reduced form of the reality.

As an alternative we can view the reduced form of the reality as
encapsulating the ``physical meaning'' of the universe. In this
alternative $u$ does not have any physical meaning. It is only the
relationships among the inferences about $u$ that one might want to
make and the devices with which to try to make those inferences that
has physical meaning. One could completely change the space $U$ and
the functions defined over it, but if the associated reduced form of
the reality does not change, then there is no way that the devices in
that reality, when considering the functions in that reality, can tell
that they are now defined over a different $U$.  In this view, the
laws of physics i.e., a choice for the set $U$, are simply a
calculational shortcut for encapsulating patterns in the reduced form
of the reality. It is a particular instantiation of those patterns
that has physical meaning, not some particular element $u \in U$.

Given a reality $(U; \{(X_1, Y_1), (X_2, Y_2),
\ldots \})$, we say that a pair of devices in it are
{\bf{pairwise}} distinguishable if they are distinguishable. We say
that a device $(X_i, Y_i)$ in that reality is {\bf{outside
distinguishable}} iff $\forall \; x_i \in X_i(U)$ and all $x'_{-i}$ in
the range of $\bigotimes_{j \ne i} X_j$, there is a $u
\in U$ such that simultaneously $X_i(u) = x_i$ and $X_j(u) = x'_j \;
\forall j \ne i$. (Note that that range may be a proper subset
of $\varprod_{j \ne i} X_j(U)$.)
We say that the reality as a whole is {\bf{mutually (setup)
distinguishable}} iff $\forall \; x_1 \in X_1(U), x_2 \in X_2(U),
\ldots \; \exists \; u
\in U$ s.t. $X_1(u) = x_1, X_2(u) = x_2, \ldots$. 

$ $

\noindent {\bf{Proposition 3:}} 

{\bf{i)}} There exist realities $(U; C_1, C_2, C_3)$ where each pair
of devices is setup distinguishable

$\;\;\;\;$and $C_1 > C_2 > C_3 > C_1$. 

{\bf{ii)}} There exists no reality $(U; \{C_i : i \in {\mathscr{N}}
\subseteq {\mathbb{N}}\})$ where the devices are mutually

$\;\;\;\;$distinguishable and for some integer $n$, $C_1 > C_2 >
\ldots > C_n > C_1$.

{\bf{iii)}} There exists no reality $(U; \{C_i : i  \in {\mathscr{N}}
\subseteq {\mathbb{N}}\})$ where for some
integer $n$, $C_1 \gg C_2 \gg$

$\;\;\;\;\ldots \gg C_n \gg C_1$.

$ $

Consider a reality with a countable set of devices $\{C_i\}$. There
are many ways to view such a reality as a graph, for example by having
each node be a device while the edges between the nodes concern
distinguishability of the associated devices, or concern whether one
weakly infers the other, etc. There are restrictions on what graphs of
those various sorts can exist.  As an example, given a countable
reality, define an associated directed graph by identifying each
device with a separate node in the graph, and by identifying each
relationship of the form $C_i \gg C_j$ with a directed edge going from
node $i$ to node $j$.  We call this the {\bf{strong inference graph}}
of the reality.

Thm. 3 means that a universal device in a reality must be a root node
of the strong inference graph of the reality. Applying Th. 3 again
shows that the strong inference graph of a reality with a universal
device must contain exactly one root.  In addition, by Thm. 2(ii), we
know that every node in a reality's strong inference graph has edges
that lead directly to every one of its successor nodes (whether or not
there is a universal device in the reality). By Prop. 3(iii) we also
know that a reality's strong inference graph is acyclic. This latter
fact establishes the following:

$ $

\noindent {\bf{Proposition 4:}} Let $D$ be a finite subset of the
devices in a reality, where the strong inference graph of the reality
is weakly connected over $D$. Say that any pair of distinct devices in
$D$ that are not connected by an edge of the strong inference graph
are setup distinguishable.  

Then the strong inference graph of the reality has one and only one
root over $D$.

$ $

\noindent Results of this sort mean there are unavoidable asymmetries in the
strong inference graphs of realities. These asymmetries provide a
preferred direction of strong inference in realities, akin to the
preferred direction in time provided by the second law of
thermodynamics.

Note that even if a device $C_1$ can strongly infer all other devices
$C_{i > 1}$ in a reality, it may not be able to infer them
$simultaneously$ (strongly or weakly). For example, define $\Gamma : u
\rightarrow (Y_2(u), Y_3(u), \ldots)$. Then the fact that $C_1$ is a
universal device does not mean that $\forall f \in \pi(\Gamma) \;\exists
\; x_1 : Y_1 = f(\Gamma)$. See the discussion in~\cite{wolp01} on
``omniscient devices'' for more on this point.

We now define what it means for two devices to operate in an identical
manner:

$ $

\noindent {\bf{Definition 7:}} Let $U$ and $\hat{U}$ be two (perhaps
identical) sets. Let $C_1$ be a device in a reality with domain
$U$. Let $R_1$ be the relation between $X_1$ and $Y_1$ specified by
the reduced form of that reality, i.e., $x_1 R_1 y_1$ iff the pair
$(x_1, y_1)$ occurs in some tuple in the reduced form of the
reality. Similarly let $R_2$ be the relation between $X_2$ and $Y_2$
for some separate device $C_2$ in the reduced form of a reality having
domain ${\hat{U}}$.

Then we say that $C_1$ {\bf{mimics}} $C_2$ iff there is an injection,
$\rho_X : X_2({\hat{U}}) \rightarrow X_1(U)$ and a bijection $\rho_Y :
Y_2({\hat{U}}) \leftrightarrow Y_1(U)$, such that for $\forall x_2,
y_2$, $x_2 R_2 y_2 \Leftrightarrow \rho_X(x_2) R_1 \rho_Y(y_2)$. If
both $C_1$ mimics $C_2$ and vice-versa, we say that $C_1$ and $C_2$
are {\bf{copies}} of each other.

$ $
  
\noindent Note that because $\rho_X$ in Def. 7 may not be surjective,
one device may mimic multiple other devices. (Surjectivity of $\rho_Y$
simply reflects the fact that since we're considering devices, $Y_1(U)
= Y_2(U) = {\mathbb{B}}$.) The relation of one device mimicing another
is reflexive and transitive. The relation of two devices being copies
is an equivalence relation.

Intuitively, when expressed as devices, two physical systems are
copies if they follow the same inference algorithm with $\rho_X$ and
$\rho_Y$ translating between those systems. In particular, say a
reality contains two separate physical computers that are inference
devices, both being used for prediction. If those devices are copies
of each other, then they form the same conclusion for the same value
of their setup function, i.e., they perform the same computation for
the same input.

As another example, say that the states of some physical system $S$ at
a particular time $t$ and shortly thereafter at $t + \delta$ are
identified as the setup and conclusion values of a device $C_1$. In
other words, $C_1$ is given by the functions $(X_1(u), Y_1(u))
\triangleq (S(u_t), S(u_{t+\delta}))$. In addition, let $R_S$ be the
relation between $X_1$ and $Y_1$ specified by the reduced form of the
reality containing the system. Say that the time-translation of $C_1$,
given by the two functions $S(u_{t'})$ and $S(u_{t' +
\delta})$, also obeys the relation $R_S$. Then the pair of functions
$(X_2(u), Y_2(u)) \triangleq (S(u_{t'}), S(u_{t' + \delta}))$ is
another device that is copy of $C_1$. So for example, the same
physical computer at two separate pairs of moments is two separate
devices, devices that are copies of each other, assuming they have the
same set of allowed computations.

Say that an inference device $C_2$ is being used for observation and
$C_1$ mimics $C_2$. The fact that $C_1$ mimics $C_2$ does not imply
that $C_1$ can emulate the observation that $C_2$ makes of some
outside function $\Gamma$. The mimicry property only relates $C_1$ and
$C_2$, with no concern for third relationships with any third
function. (This is why for one device to ``emulate'' another is
defined in terms of strong inference rather than in terms of mimicry.)

Next for future use we note the following fact that is almost obvious
(despite being so complicated):

$ $

\noindent{\bf{Lemma 1:}} Let $K_1$ be the set of reduced forms of all device
realities. Let $K_2$ be the set of all sets $k$ with the following
property: $k$ can be written as $\{(\varprod_{\alpha \in
{\mathscr{A}}} (s^r_\alpha, t^r_\alpha) \times \varprod_{\beta \in
{\mathscr{B}}} v^r_\beta) : r
\in R\}$ for some associated ${\mathscr{A}}, {\mathscr{B}}$ and $R$ such
that for all $\alpha$, $\cup_r t^r_\alpha = {\mathbb{B}}$ and $|\cup_r
s^r_\alpha| \ge 2$, while for all $\beta \in {\mathscr{B}}$, $|\cup_r
v^r_\beta| \ge 2$. Then $K_1 = K_2$. In particular, any $k \in K_2$ is
the reduced form of a reality $(U;
\{C_\alpha\}, \{\Gamma_\beta\})$, where for all $\alpha \in
{\mathscr{A}}, \beta \in {\mathscr{B}}, u \in U$, there is some
associated $r \in R$ such that simultaneously $X_\alpha(u) =
s^r_\alpha, Y_\alpha(u) = t^r_\alpha$, and $\Gamma_\beta(u) =
v^r_\beta$.

$ $

\noindent Next, fix a counting number $m$ and a set of $m$
cardinalities, $\{\Omega_i : i = 1, \ldots m\}$.  Let $M$ be the set
of all realities each of which comprises $m$ functions, where the
ranges of those $m$ functions have the associated cardinalities
$\{\Omega_i : i = 1, \ldots m\}$.

Now say we ask whether there is a reality in $M$ whose $m$ functions
have some particular relationship(s) with one another. (Answers to
such questions form most of the results of the earlier parts of this
paper.)  Lemma 1 allows us to transform this question into a
constraint satisfaction problem over an associated space of
tuples. This transformation changes set of ``specified
relationship(s)'' into a set of simultaneous constraints over the
associated space of tuples. The precise type of constraint
satisfaction problem produced by the transformation (integer-valued,
real-valued, etc.) is determined by the space of tuples under
consideration, i.e., by the cardinalities of the images of the
functions that constitute the reality.

Often though we can use Lemma 1 more directly to answer questions
concerning realities, without invoking any techniques for solving
constraint satisfaction problems. An example occurs in the proof of
the following result:

$ $
  
\noindent {\bf{Proposition 5:}} Let $C_1$ be a copy of $C_2$.

{\bf{i)}} It is possible that $C_1$ and $C_2$ are distinguishable and
$C_1 > C_2$, even for finite $X_1(U), X_2(U)$.

{\bf{ii)}} It is possible that $C_1 \gg C_2$, but only if $X_1(U)$ and
$X_2(U)$ are both infinite.

\subsection{Philosophical implications}
\label{sec:philo}

Return now to the case where $U$ is a set of laws of physics (i.e.,
the set of all worldlines consistent with a set of such laws). The
results of this subsection provide general restrictions that must
relate any devices in such a universe, regardless of the detailed
nature of the laws of that universe. In particular, these results
would have to be obeyed by all universes in a
multiverse~\cite{smol02,agte05,carr07}.

Accordingly, it is interesting to consider these results from an
informal philosophical perspective. Say we have a device $C$ in a
reality that is outside distinguishable. Such a device can be viewed
as having ``free will'', in that the way the other devices are set up
does not restrict how $C$ can be set up.  Under this interpretation,
Thm. 1 means that if two devices both have free will, then they cannot
predict / recall / observe each other with guaranteed complete
accuracy.  A reality can have at most one of its devices that has free
will and can predict / recall / observe the other devices in that
reality with guaranteed complete accuracy. (Similar conclusions hold
for whether the devices can ``control'' each other; see
Sec.~\ref{sec:control} below.)

Thm. 3 then goes further and considers devices that can emulate each
other. It shows that independent of concerns of free will, no two
devices can unerringly emulate each other. (In other words, no reality
can have more than one universal device.) Somewhat tongue in cheek,
taken together, these results could be called a ``monotheism
theorem''.

Now suppose that the domain of a reality is a set of worldlines
extending across time, and consider ``physical'' devices that are
identified with systems evolving in time. (See discussion just after
Def. 7.) Prop. 5 tells us that any universal device must be infinite
(have infinite $X(U)$) if there are other devices in the reality that
are copies of it. Since the time-translation of a physical device is a
copy of that device, this means any physical device that is ever
universal must be infinite. In addition, the impossibility of multiple
universal devices in a reality means that if any physical device is
universal, it can only be so at one moment in time. (Its
time-translation cannot be universal.) Again somewhat tongue in cheek,
taken together this second set of results could be called an
``intelligent design theorem''. (See Sec.~\ref{sec:control} for
related limitations concerning devices that are used to control one
another.)

In addition to the questions addressed by the monotheism and
intelligent design theorems, there are many other semi-philosophical
questions one can ask of the form ``Can there be a reality with the
following properties?''. As mentioned above, Lemma 1 can be used to
reduce all such questions to a constraint satisfaction problem,
potentially involving infinite-dimensional spaces. In other words,
much of philosophy can be reduced to constraint satisfaction
problems.

As a final comment, while it is most straight-forward to apply the
results of this subsection to physical universes, they can be applied
more widely. In particular, somewhat speculatively, one can consider
applying them to mathematical logic itself. In such an application
each $u \in U$ would be a (perhaps infinite) string over some
alphabet. For example, $U$ might be defined as the set of all strings
that are ``true'' under some encoding that translates a string into
axioms and associated logical implications. Then an inference device
would be a (perhaps fallible) theorem-proving algorithm, embodied
within $U$ itself. The results of this subsection would then concern
the relation among such theorem-proving algorithms.

\section{Control devices}
\label{sec:control}

In weak inference there is no causal arrow from $\Gamma$ to $X$. In
fact, the only causal arrow goes from the device to the function being
inferred (in that $X$'s value forces something about $\Gamma$'s value)
rather than vice-versa. This reflects what it means for us to be able
to set up a device so that it is guaranteed correct in its prediction
/ observation/ memory.

This causal arrow from the device to the function does not mean that
the device controls the function. The reason is that $X$'s value
doesn't set $\Gamma$'s value, but only forces that value to be
consistent with $Y$. This motivates the following definition:

$ $

\noindent {\bf{Definition 8:}} A device $C$ {\bf{controls}} a
function $\Gamma$ over ${U}$ iff $\forall \; f \in \pi(\Gamma)$,
$\forall b \in {\mathbb{B}}, \exists x$ such that $X = x
\Rightarrow Y = f({\Gamma}) = b$. $C$ {\bf{semi-controls}}
$\Gamma$ iff $\forall \gamma \in \Gamma(U)$, $\exists\; x$ such that
$X=x \Rightarrow \Gamma = \gamma$.

$ $

Semi-control has nothing to do with the conclusion function $Y$ of the
device; that function enters when one strengthens the definition of
semi-control to get the definition of control. To see this, note that
$C$ semi-controls $\Gamma$ iff $\forall \; f
\in \pi(\Gamma)$, $\exists x$ such that $X = x
\Rightarrow f({\Gamma}) = 1$. However if $X=x$ forces $f(\Gamma) = 1$, then for
any probe $f' \ne f$, $X = x$ forces $f'(\Gamma) = 0$. So $C$
semi-controls $\Gamma$ iff $\forall \; f
\in \pi(\Gamma)$, $\forall b
\in {\mathbb{B}}, \exists x$ such that $X = x
\Rightarrow f({\Gamma}) = b$. This is just the definition of control, without the extra
condition that controls imposes on the value of $Y$.  We say that one
device $C$ (semi-) controls another if it (semi-) controls the
conclusion function of that second device.

The weakness of the semi-control concept is that it stipulates nothing
concerning whether $C$ ``knows'' (infers) that some value $x$ forces
$\Gamma$ into the state $f^{-1}(b)$.  In this, it doesn't capture the
intuitive notion of ``control''. Accordingly, in the formalization of
Def. 8, we stipulate that you do not fully control a function if you
force it to have some value but don't know what that value is.

If the partition induced by $X$ is a refinement of the partition
induced by $\Gamma$~\cite{albo06}, and in particular if it is a
fine-graining of that partition, then $C$ semi-controls $\Gamma$.
Note also that if $\Gamma$ is binary-valued, then having $C$
semi-control $\Gamma$ means there is both an $x$ such that $X(u) = x
\Rightarrow u \in \Gamma^{-1}(1)$ and an $x'$ such that $X(u) = x'
\Rightarrow u \in \Gamma^{-1}(-1)$. In the language of formal
epistemology~\cite{auma99,aubr95,futi91,bibr88}, this means that
$X^{-1}(x)$ and $X^{-1}(x')$ are the values of a ``knowledge
function'' evaluated for two arguments: the subset
$\Gamma^{-1}(1)$ and the subset $\Gamma^{-1}(-1)$, respectively. (See
Sec.~\ref{sec:sad} below.)

Clearly control implies semi-control. In addition, if one device $C_1$
strongly infers another device $C_2$, then $C_1$ semi-controls $X_2$,
though it may not semi-control $Y_2$. Control implies weak inference,
i.e., if $C_1$ controls a function $\Gamma$ then $C_1 > \Gamma$. The
logical converse need not hold though.

Since control implies weak inference, all impossibility results
concerning weak inference also apply to control. In particular, no
device can control itself, and no two distinguishable devices can
control each other. In fact we can make the following stronger
statement, which essentially states that if two partitions are
refinements of each another, they must be identical:

$ $

\noindent {\bf{Theorem 5:}} If two devices $C_1$ and
$C_2$ simultaneously semi-control one another's setup functions, then
the partitions induced by $X_1$ and $X_2$ are identical.

$ $

\noindent Intuitively, Thm. 5 means that if two devices simultaneously
semi-control one another's setup functions, then those setup functions
are identical, up to a relabeling of their ranges. This provides the
following results contrasting with Thm. 1 and Thm. 3:

$ $

\noindent {\bf{Corollary 3:}} Let $C_1$ and
$C_2$ be two devices that simultaneously semi-control one another's
setup functions. 

{\bf{i)}} $C_1 > C_2 \Leftrightarrow C_2 > C_1$.

{\bf{ii)}} Neither device strongly infers the other.

{\bf{iii)}} Neither device controls the other's setup function.

$ $

\section{Stochastic devices}

In the analysis above there is no probability measure $P$ over $U$.
There are several ways to extend the analysis to incorporate such a
probability measure, so that functions over $U$ become random
variables. One starts as follows:

$ $

\noindent {\bf{Definition 9:}} Let $P(u
\in U)$ be a  probability measure,  $\Gamma$ a function with
domain $U$ and finite range, and $\epsilon
\in [0.0, 1.0]$. Then we say that a device $(X, Y)$
(weakly) infers $\Gamma$ {\bf{with (covariance) accuracy}} $\epsilon$ iff
\begin{eqnarray*}
\frac{\sum_{f \in \pi(\Gamma)}{\mbox{max}}_{x} [{\mathbb{E}}_P(Y f(\Gamma) \mid x)]}{|(\Gamma(U)|} &=&
\epsilon.
\end{eqnarray*}
\noindent  As an example, if $P$ is nowhere 0 and
$C$ weakly infers $\Gamma$, then $C$ infers $\Gamma$ with accuracy 1.0.{\footnote{A
subtlety with the definition of an inference devices arises in this stochastic setting:
we can either require that $Y$ be surjective, as in Def. 1, or instead require that
$Y$ be {\bf{stochastically surjective}}: $\forall y \in {\mathbb{B}}, \; \exists
u$ with non-zero probability density such that $Y(u) = y$. The distinction
between requiring surjectivity and stochastic surjectivity of $Y$ will not
arise here.}}

There are several reasonable alternatives to this definition. As an
example, recall the ``malicious demon'' interpretation of $f$
introduced just below Def. 3. That interpretation suggests a change to
Def. 9 in which we replace the sum over all probes $f$ and associated
division by $|\Gamma(U)|$ with a minimum over all probes $f$.  

Note though that it does $not$ seem reasonable to define inference
accuracy in terms of mutual information expressions like
${\mathbb{M}}(Y, f(\Gamma) \mid X = x)$. To see why consider the case
where $f$ is a probe of $\Gamma$ that equals 1 iff $\Gamma = \gamma$,
and let $x$ be a value where $X = x \Rightarrow Y = -f(\Gamma)$. In
this case the mutual information conditioned on $x$ between $Y$ and
$f(\Gamma)$ would be maximal. However the device would have
probability zero of correctly answering the question, ``does $\Gamma$
have value $\gamma$?''. It would either say ``yes'' and in fact
$\Gamma$ does not equal $\gamma$, or it would say ``no'' and in fact
$\Gamma$ does equal $\gamma$.

This is an illustration of the fact that the definition of inference
assigns semantic content to $Y = 1$: it means that the device's answer
is ``yes''. In contrast, information theoretic quantities like mutual
information are (in)famous for not involving semantic content.

While inference is a semantic concept, distinguishability is not,
which motivates the following definition:

$ $

\noindent {\bf{Definition 10:}} Let $P(u
\in U)$ be a  probability measure, and $\epsilon
\in [0.0, 1.0]$. Then we say that the {\bf{(setup) mutual
information-distinguishability}} of two device $(X_1, Y_1)$ and $(X_2,
Y_2)$ is 
\begin{eqnarray*}
1 - \frac{{\mathbb{M}}_P(X_1, X_2)}{{\mathbb{H}}_P(X_1) +
{\mathbb{H}}_P(X_2)}.
\end{eqnarray*}

$ $

\noindent Mutual-information distinguishability  is bounded between
0 and 1.

Note that variables can be distinguishable in the sense of Def. 4 even
if their mutual information distinguishability is less than 1. (They
can be partially correlated but still distinguishable in the sense of
Def. 4.) This motivates the following alternative definition, for
simplicity phrased for countable $X(U)$:

$ $

\noindent {\bf{Definition 11:}} Let $P(u
\in U)$ be a  probability measure, and $\epsilon
\in [0.0, 1.0]$. Then we say that the {\bf{counting
distinguishability}} of two device $(X_1, Y_1)$ and $(X_2,
Y_2)$ is 
\begin{eqnarray*}
1 - \frac{\sum_{x_1, x_2 \; : \; \exists \; u \; : \;
X_1(u) = x_1, X_2(u) = x_2} 1}{|X_1(U)| \times |X_2(U)|}
\end{eqnarray*}
$ $

There are many analogs of Thm. 1 that relate quantities like the
accuracy with which device $C_1$ infers device $C_2$, the accuracy
with which $C_2$ infers $C_1$, how distinguishable they are, the
entropies of the random variables $X_1$ and $X_2$, etc.  To present
perhaps the simplest such example, define $H$ as the four-dimensional
hypercube $\{0, 1\})^4$, $k({\vec{z}})$ as the map taking any
${\vec{z}} \in H$ to $z_1 + z_4 - z_2 - z_3$, $m({\vec{z}})$ as the
map taking any ${\vec{z}} \in H$ to $(z_2 - z_4)$, and $n({\vec{z}})$
as the map taking any ${\vec{z}} \in H$ to $(z_3 - z_4)$.

$ $

\noindent {\bf{Proposition 6:}} Let $P$ be a probability measure over $U$,
and $C_1$ and $C_2$ two devices whose mutual-information
distinguishability is 1, where $X_1(U) = X_2(U) = {\mathbb{B}}$. Define
$P(X_1 = -1) \equiv \alpha$ and $P(X_2 = -1)
\equiv
\beta$. Say that $C_1$ infers $C_2$ with accuracy $\epsilon_1$, while
$C_2$ infers $C_2$ with accuracy $\epsilon_2$. Then
\begin{eqnarray*}
\epsilon_1 \epsilon_2 \;&\le&\; {\mbox{max}}_{{\vec{z}} \in H} \;
| \; \alpha \beta [k({\vec{z}})]^2 + \alpha k({\vec{z}})m({\vec{z}}) +
\beta k({\vec{z}})n({\vec{z}}) + m({\vec{z}})n({\vec{z}})\; | .
\end{eqnarray*}
In particular, if $\alpha = \beta = 1/2$, then
\begin{eqnarray*}
\epsilon_1 \epsilon_2 \;&\le&\; \frac{{\mbox{max}}_{{\vec{z}} \in H}
\; |\; (z_1 -
z_4)^2 - (z_2 - z_3)^2\; |}{4} \nonumber \\
&=&\; 1/4.
\end{eqnarray*}

$ $

\noindent The maximum for $\alpha = \beta = 1/2$ can occur in several
ways. One is when $z_1 = 1$, and $z_2, z_3, z_4$ all equal $0$. At
these values, both devices have an inference accuracy of 1/2 at
inferring each other. Each device achieves that accuracy by perfectly
inferring one probe of the other device, while performing randomly for
the remaining probe.

Similarly, say that we have a volume measure $d\mu$ over $U$, as in
Sec.~\ref{sec:inf_compl}, together with a probability measure $P$ over
$U$. Then we can modify the definition of the length of $x$ to be
$-{\mathbb{H}}(U \mid x)$, the negative of the Shannon entropy under
prior $d\mu$ of $P(u \mid x)$.  If as in statistical physics $P$ is
proportional to $d\mu$ across the support of $P$, then $P(u \mid x)
\propto d\mu(u
\mid x)$, and these two definitions of the length of $x$ are the same.

There are several ways to combine this new definition of length with
the concept of inference accuracy to define a stochastic analog of
inference complexity. In particular, we can define the {\bf{stochastic
inference complexity}} of a function $\Gamma$ with respect to $C$ for
accuracy $\epsilon$, as
\begin{eqnarray*}
{\bar{\mathscr{C}}}_\epsilon(\Gamma \mid C) \;\;&\triangleq& \;\; \sum_{f \in \pi(\Gamma)}
	{\mbox{min}}_{x : {\mathbb{E}}_P(Y f(\Gamma) \mid x) \ge \epsilon}
			 [-{\mathbb{H}}(U \mid x)]
\end{eqnarray*}
assuming the sum exists for $\epsilon$.  So for example if $P$ is
proportional to $d\mu$ across the support of $P$ and $C > \Gamma$,
then for $\epsilon = 1$, ${\bar{\mathscr{C}}}_\epsilon(\Gamma \mid C) =
{\mathscr{C}}(\Gamma \mid C)$.

One can extend this stochastic framework to include inference of the
probability of an event, e.g., have the device say whether $P(\Gamma =
\gamma)$ has some specified value. Such inference contrasts with
inference accuracy, which (like non-stochastic inference) simply
concerns a device's concluding whether an event occurs, e.g.,
concluding whether $\Gamma(u) = \gamma$). One can also define
stochastic analogs of (semi)control, strong inference, etc. Such
extensions are beyond the scope of this paper.

\section{Self-aware devices}
\label{sec:sad}

We now return to scenarios where $U$ has no associated probability
measure. We consider devices that know what question they are
trying to answer, or at least ``think they do''. Rather than encode
that knowledge in the conclusion function of the device, we split the
conclusion function into two parts. The value of one of those parts is
(explicitly) a question for the device, and the other part is a
possible associated answer. We formalize this as follows:

$ $

\noindent {\bf{Definition 12:}} A {\bf{self-aware}} device is a triple
$(X, Y, Q)$ where $(X, Y)$ is an inference device, $Q$ is a
{\bf{question}} function with domain $U$ where each $q \in Q(U)$ is a
binary function of $U$, and $Y \otimes Q$ is surjective onto
${\mathbb{B}} \times Q(U)$.

$ $

\noindent Intuitively, a self-aware device is one that (potentially) knows what
question it is answering in its conclusion. When $U =u$, we interpret
$q = Q(u)$ as the question about the state of the universe (i.e.,
about which subset of $U$ contains the actual $u$) that the conclusion
$Y(u)$ is supposed to answer. The reason we require that $Y \otimes Q$
be surjective onto ${\mathbb{B}} \times Q(U)$ is so that the device is
allowed to have any conclusion for any of its questions; it's the
appropriate setting of $X(u)$ that should determine what conclusion it
actually makes.

So one way to view ``successful inference'' is the mapping of any $q
\in Q(U)$ to an $x$ such that $X(u) = x(u)$ both implies that the device's  conclusion
to question $q$ is correct, i.e., $Y(u) = q(u)$, and also implies that
the device is sure it is asking question $q$, i.e., $Q(u) = q$. As an
example, say we have a computer that we want to use make a
prediction. That computer can be viewed as an inference device. In
this case the question $q$ that the device is addressing is specified
in the mind of the external scientist.  This means that the question
is a function of $u$ (since the scientist exists in the universe), but
need not be stored directly in the inference device. Accordingly, the
combination of the computer with the external scientist who programs
the computer is a self-aware device.

To formalize this concept, we must first introduce some notation that
is frankly cumbersome, but necessary for complete precision. Let $b$
be a value in some space. Then we define ${\underline{b}}$ as the
constant function over $U$ whose value is $b$, i.e., $u \in U
\rightarrow b$. Intuitively, the underline operator takes any constant
and produces an associated constant-valued function over $U$.  As a
particular example, let $\Gamma$ be a function with domain $U$. Then
${\underline{\Gamma}}$ is the constant function over $U$ whose value
is the function $\Gamma$, i.e., $u \in U \rightarrow
\Gamma$. Similarly, let $B$ be a set of functions with domain $U$, and
let $A$ be a function with domain $U$ whose range is $B$ (so each
$A(u)$ is a function over $U$). Then we define ${\overline{A}}$ as the
function taking $u \in U \rightarrow [A(u)](u)$. So the overline
operator turns any function over $U$ whose range is functions over $U$
into a single function over $U$. Both the underline and overline
operators turn mathematical structures into functions over $U$; they
differ in what type of argument they take. In particular, for any
function $\Gamma$ over $U$, ${\overline{(\underline{\Gamma})}} =
\Gamma$. (Using this notation is more intuitive in practice than
these complicated definitions might suggest.)

Next, recall from Sec.~\ref{sec:notation} that for any probe $f$ of a
function $\Gamma$ with domain $U$, $f(\Gamma)$ is the function $u \in
U \rightarrow f(\Gamma(u))$. 

$ $

\noindent {\bf{Definition 13:}} Let $D = (X, Y, Q)$ be a self-aware
device.

{\bf{i)}} A function $\Gamma$ is {\bf{intelligible}} to $D$ iff
$\forall \; f \in \pi(\Gamma)$, $f(\Gamma) \in Q(U)$.

{\bf{ii)}} $D$ is {\bf{infallible}}  iff $\forall u \in U$, $Y(u) =
[Q(u)](u)$.

$ $

\noindent  We say that $D$ is infallible for $Q' \subseteq Q(U)$ iff
$\forall q \in Q'$, $\forall u \in U$ such that $Q(u) = q$, $Y(u) =
q(u)$. So $D$ is infallible iff it is infallible for $Q(U)$ iff $Y =
{\overline{Q}}$ iff $Y {\overline{Q}} = {\underline{1}}$. If a device
is not infallible, we say that it is fallible.

Recall that $Y \otimes Q$ is supposed to represent the original
conclusion function ``split into two parts''. Accordingly, in keeping
with the terminology used with weak inference, we say that a
self-aware device $(X', Y', Q')$ is intelligible to a self-aware
device $(X, Y, Q)$ iff $(Y', Q')$ is intelligible to $(X, Y, Q)$.

Def. 13 provides the extra concepts needed to analyze inference with
self-aware devices. Def. 13(i) means that $D$ is able to ask what the
value is of every probe of $\Gamma$. Def. 13(ii) ensures that
$whatever$ the question $D$ is asking, it is correctly answering that
question. Finally, the third part of ``successful inference'' ---
having the device be sure it is asking the question $q$ --- arises if
$D$ semi-controls its question function.

These definitions are related to inference by the following results:

$ $

\noindent {\bf{Theorem 6:}} Let $D_1$ be an infallible,
self-aware device.

{\bf{i)}} Let $\Gamma$ be a function intelligible to $D_1$ and say
that $D_1$ semi-controls $Q_1$. Then $(X_1, Y_1) > \Gamma$.

{\bf{ii)}} Let $D_2$ be a device where $Y_2$ is intelligible to $D_1$,
$D_1$ semi-controls $(Q_1, X_2)$, and $(Q_1, X_2)$

$\;\;\;\;$is
surjective onto $Q_1(U) \times X_2(U)$. Then $(X_1, Y_1) \gg (X_2, Y_2)$.

$ $

\noindent  Thm. 6 allows us to apply results concerning weak
and strong inference to self-aware devices. Note that a special case
of having $D_1$ semi-control $Q_1$ is where $X = \chi \otimes Q_1$ for
some function $\chi$, as in Ex. 1. For such a case, $Y$ and $X$
``share a component'', namely the question being asked, specified in
$Q_1$.

The following result concerns just intelligibility, without
any concern for semi-control or infallibility. 

$ $

\noindent {\bf{Theorem 7:}} Consider a pair of self-aware devices
$D \equiv (X, Y, Q)$ and $D' \equiv (X', Y', Q')$ where there are
functions $R, P, R', P'$ such that $P$ and $P'$ have domain $U$, $Q =
R(P)$ and $Q' = R'(P')$. If $P$ is intelligible to $D'$ and $P$ is
intelligible to $D'$ then the following hold:

{\bf{i)}} $|Q(U)| = |Q'(U)| = |P(U)| = |P'(U)|$. 

{\bf{ii)}} If $Q(U)$ is finite, $Q' = \pi(P) = \pi(Q)$ and $Q =
\pi(P') = \pi(Q')$.

$ $

\noindent In particular, take $R$ and $R'$ to be identity functions
over the associated domains, so that $P = Q$ and $P' = Q'$. Using this
choice, Thm. 7 says that if each self-aware device can try to
determine what question the other one is considering, then neither
device can try to determine anything else.

An immediate corollary  of Thm. 7 is the following:

$ $

\noindent {\bf{Corollary 4:}} No two self-aware devices whose
question functions have finite ranges are intelligible to each other.

$ $

\noindent Note that Coroll. 4 does not rely on the devices being
distinguishable (unlike Thm. 1). Indeed, it holds even if the two
devices are identical; a self-aware device whose question function has
a finite range cannot be intelligible to itself.

Coroll. 4 is a powerful limitation on any pair of self-aware
devices, $D$ and $D'$. It says that for at least one of the devices,
say $D$, there is some question $q'
\in Q'(U)$ and bit $b'$, such that $D$ cannot even $ask$, ``Does $D'$
pose the question $q'$ and answer with the bit $b'$?''. So whether $D$
could correctly answer such a question is moot.

To circumvent Coroll. 4 we can consider self-aware devices whose
conclusion functions alone are intelligible to each other. However
combining Thm.'s 1 and 3(i) gives the following result:

$ $

\noindent {\bf{Corollary 5:}} Let $D_1$ and $D_2$ be  two  
self-aware devices that are infallible, semi-control their questions,
and are distinguishable. If in addition they infer each other, then it
is not possible that both $Y_2$ is intelligible to $D_1$ and $Y_1$ is
intelligible to $D_2$.

$ $

With self-aware devices a device $C_1$ may be able to infer whether a
self-aware device $D_2$ correctly answers the question that $D_2$ is
considering. To analyze this issue we start the following definition:

$ $ 

\noindent {\bf{Definition 14:}} If $D_1$ is a device and $D_2$ a
self-aware device, then $D_1$ {\bf{corrects}} $D_2$ iff $\exists\;
x_1$ such that $X_1 = x_1 \Rightarrow Y_1 = Y_2 {\overline{Q_2}}$.

$ $

\noindent Def. 2  means that $Y_1 = 1$ iff $Y_2 =
{\overline{Q_2}}$, i.e., $Y_2(u) = [Q_2(u)](u)$. Intuitively, if a
device $D_1$ corrects $D_2$, then there is an $x_1$ where having $X_1
= x_1$ means that $C_1$'s conclusion tells us whether $D_2$ correctly
answers $q_2$.{\footnote{\label{foot_1} Say that $D_1$ is also
self-aware, and that $Y_2 {\overline{Q_2}}$ has both bits in its range
(so that probes of it are well-defined). Then we can modify the
definition to say that $D_1$ corrects $D_2$ iff two conditions are
met: all probes in $\pi(Y_2 {\overline{Q_2}})$ are intelligible to
$D_1$, and $D_1$ is infallible for $\pi(Y_2 {\overline{Q_2}})$.}}

Note how weak Def. 14 is. In particular, there is no sense in which
it requires that $D_1$ can assess whether $Y_2(u) = q_2(u)$ for all
questions $q_2 \in Q_2(U)$. So long as $D_1$ can make that assessment
for $any$ question in $Q_2(U)$, we say that $D_1$ corrects
$D_2$. Despite this weakness, we have the following impossibility
result, which is similar to Prop. 2(i):

$ $

\noindent {\bf{Proposition 7:}} For any device $D_1$ there is a
self-aware device $D_2$ that $D_1$ does not correct. 

$ $

\noindent 
There are similar results for the definition of correction in
footnote~\ref{foot_1}, and for the (im)possibility of correction among
multiple devices.

Finally, while there is not room to do so here, many of the concepts
investigated above for inference devices can be extended to self-aware
devices. For example, one might want to modify the definition of
inference complexity slightly for self-aware devices. Let $D$ be a
self-aware infallible device that semi-controls its question function
and $\Gamma$ a function over $U$ where $\Gamma(U)$ is countable and
$\Gamma$ is intelligible to $D$. Then rather than
${\mathscr{C}}(\Gamma \mid (X, Y))$, it may be more appropriate to
consider the {\bf{self-aware inference complexity}} of $\Gamma$ with
respect to $D$, defined as
\begin{eqnarray*}
{\mathscr{D}}(\Gamma \mid (X,Y,Q)) \;\;&\triangleq& \;\; \sum_{f \in \pi(\Gamma)}
	{\mbox{min}}_{x : X = x \Rightarrow Q = {\underline{f(\Gamma)}}}
			 [{\mathscr{L}} (x)].
\end{eqnarray*}

\noindent Similarly, consider a reality that includes self-aware
devices, i.e., a reality $(U; \{F_\phi\})$ that can be written as $(U;
\{C_\alpha\}; \{D_\delta\}; \{\Gamma_\beta\})$ where in addition to the set of
functions $\{\Gamma_\beta\}$ and devices $\{C_\alpha\}$, we have a set
of self-aware devices $\{D_\delta\}$. For such a reality it often
makes sense to consider an augmented reduced form, 
\begin{eqnarray*}
\bigcup_{u \in U}
\left[ \bigotimes_\alpha (X_\alpha(u), Y_\alpha(u)) \otimes
\bigotimes_\beta \Gamma_\beta(u) \otimes \bigotimes_\delta (X_\delta(u), Y_\delta(u),
Q_\delta(u)) \otimes \bigotimes_\delta Q_\delta(U) \right].
\end{eqnarray*}
The last term means we include in the tuples all instances of the form
$[Q(u)](u')$ in which a self-aware device's question for one $u$ is
evaluated at a different $u' \ne u$.

Due to page limits the analysis of such extensions is beyond the scope
of this paper.

We close with some comments on the relation between inference with
self-aware devices and work in other fields.  Loosely speaking, in the
many-worlds interpretation of quantum mechanics~\cite{ever57},
``observation'' only involves the relationship between $Y$ and
$\Gamma$ (in general, for a $Y$ whose range is more than binary). As
discussed above, such relationships cannot imbue the observation with
semantic meaning. It is by introducing $X$ and $Q$ into the definition
of self-aware devices that we allow an act of ``observation'' to have
semantic meaning. This is formalized in Thm. 6, when it is applied to
scenarios where weak inference is interpreted as successful
observation.

Much of formal epistemology concerns ``knowledge functions'' which are
maps from subsets of $U$ to other subsets of
$U$~\cite{auma99,aubr95,futi91,bibr88}. 
$K_i(A)$, the knowledge function $K_i$ evaluated for an argument $A
\subseteq U$, is interpreted as the set of possible worlds in which
individual $i$ knows that $A$ is true. The set $A$ is analogous to
specification of the question being asked by a self-aware device. So
by requiring the specification of $A$, knowledge functions involve
semantic meaning, in contrast to the process of observation in the
many-worlds interpretation.

A major distinction between inference devices and both the
theory of knowledge functions and the many-worlds definition of
observation is that inference devices require that the individual /
observer be able to answer multiple questions (one for each probe
concerning the function being inferred). As mentioned above, this
requirement certainly holds in all real-world instances of
``knowledge'' or ``observation''. Yet it is this seemingly innocuous
requirement that drives many of the results presented above.

Future work involves exploring what inference device theory has to say
about issues of interest in the theory of knowledge functions. For
example, analysis of common knowledge starts with a formalization of
what it means for ``individual $i$ to know that individual $j$ knows
$A$''. The inference devices analog would be a formalization of what
it means for ``device $D$ to infer that device $C$ infers $\Gamma$''.
Now for this analog to be meaningful, since $D$ can only infer
functions with at least two values in their range, there must be some
sense in which the set $U$ both contains ''$u$ under which $C$ infers
$\Gamma$'' and contains $u$ under which it does not. Formally, this
means two things. First, it must not be the case simply that $C >
\Gamma$, since that means that $C$ infers $\Gamma$ under $all$
$u$. Second, there must be a proper subset $U_C \subset U$ such that
if $U$ were redefined to be $U_C$ (and $C$ and $\Gamma$ were redefined
to have $U_C$ as their domains in the obvious way), then it $would$ be
the case that $C > \Gamma$. This proper subset specifies a
binary-valued function, $\Gamma_C$, by $\Gamma_C(u) = 1
\Leftrightarrow u \in U_C$.  The question of whether ``$D$ knows that
$C$ knows $\Gamma$'' then becomes whether $D$ can infer $\Gamma_C$.

$ $

$ $

\noindent {\bf{ACKNOWLEDGEMENTS:}} I would like to thank Nihat Ay, Charlie Bennett,
John Doyle, Michael Gogins, and Walter Read for helpful discussion.

$ $

$ $

\noindent {\bf{APPENDIX A: Proofs}}

$ $

This section presents miscellaneous proofs. Since many of the results
may be counter-intuitive, the proofs are presented in elaborate
detail. The reader should bear in mind though that many of the proofs
simply amount to ``higher order'' versions of the Cretan liar paradox,
Cantor diagonalization, or the like (just like many proofs in Turing
machine theory). At the same time, in the interest of space, little
pedagogical discussion is inserted. Unfortunately, the combination
makes many of the proofs a bit of a slog.

$ $

\noindent {\bf{Proof of Prop. 1:}} To prove (i), choose a device $(X, Y)$ where
$Y(u) = -1 \Leftrightarrow u \in W$. Also have $X(u)$ take on a
separate unique value for each $u \in W$, i.e., $\forall w \in W, u
\in U$ : $w \ne u$, $X(w) \ne X(u)$. (Note that by definition of $W$,
it contains at least two elements.) So by appropriate choice of an
$x$, $X(u) = x$ forces $u$ to be any desired element of $W$.

Choose $i$. Pick any $\gamma \in \Gamma_i(U)$, and examine the probe
$f$ that equals 1 iff its argument is $\gamma$. If for no $u \in W$
does $\Gamma_i(u) = \gamma$, then choose any $x$ that forces
$u \in W$. By construction, $X(u) = x \Rightarrow Y(u) = -1$, and in
addition $X(u) = x \Rightarrow f(\Gamma_i(u)) = -1$. So $X(u) = x
\Rightarrow Y(u) = f(\Gamma_i(u))$, as desired.

Now say that there is a $u \in W$ such that $\Gamma_i(u) =
\gamma$. By hypothesis, $\exists u'' \in W : \Gamma_i(u'') \ne
\gamma$. By construction, there is an $x$ such that $X(u') = x
\Rightarrow u' = u''.$ So $X(u') = x \Rightarrow 
u' \in W, \Gamma_i(u') \ne \gamma$. The first of those two conclusions
means that $Y(u') = -1$. The second means that $f(\Gamma_i(u')) =
-1$. So again, $X(u) = x \Rightarrow Y(u) = f(\Gamma_i(u))$, as
desired. There are no more cases to consider.

To prove (ii), choose $b \in {\mathbb{B}}$ and let $\Gamma$ be a
function with domain $U$ where $\Gamma(u) = b$ for all $u$ obeying
$Y(u) = -1$ and for no others. (The surjectivity of $Y$ ensures there
is at least one such $u$.) Consider the probe $f$ of $\Gamma(U)$ that
equals +1 iff $\Gamma(u) = b$. For all $u \in U$,
$f(\Gamma(u)) = -Y(u)$.  {\bf{QED.}}

$ $

\noindent {\bf{Proof of Coroll. 2:}} To prove the first part of the
corollary, let $\alpha$ and $\beta$ be the partitions induced by $X$
and $Y$, respectively. If $|X(U)| = |\alpha| = 2$, $|\alpha| =
|\beta|$. Since $\alpha$ is a fine-graining of $\beta$, this means
that $\alpha =
\beta$. So without loss of generality we can label the elements of
$X(U)$ so that $X = Y$.

Now hypothesize that $C > \Gamma$ for some $\Gamma$. Recall that we
require that $|\Gamma(U)| \ge 2$. Let $\gamma$ and $\gamma'$ be two
distinct elements of $\Gamma(U)$ where $\Gamma(u) = \gamma$ for some
$u \in X^{-1}(-1)$. Define $f_\gamma$ to be the probe of $\Gamma(U)$
that equals 1 iff its argument is $\gamma$, and define $f_{\gamma'}$
similarly. $C > \Gamma$ means $\exists \; x_\gamma \in X(U)$ such that
$X(u) = x_\gamma \Rightarrow f_\gamma(\Gamma(u)) = Y(u) = X(u) =
x_\gamma$. Since $\exists \; u \in X^{-1}(-1)$ such that $\Gamma(u) =
\gamma$, and since $Y(u) = -1 \; \forall u \in X^{-1}(-1)$, $x_\gamma$
must equal $1$.

This means that $\Gamma(u)$ equals $\gamma$ across all of
$X^{-1}(x_\gamma) \subset U$. Therefore $\exists \; u \in
X^{-1}(-x_\gamma)$ such that $\Gamma(u) = \gamma'$. Moreover, since
$x_\gamma = Y(X^{-1}(x_\gamma)) = 1$, $Y(X^{-1}(-x_\gamma)) =
-1$. Therefore $\exists \;u \in X^{-1}(-x_\gamma)$ such that
$f_{\gamma'}(\Gamma(u)) \ne Y(u)$. Similarly, $\forall \; u \in
X^{-1}(x_\gamma)$, $f_{\gamma'}(\Gamma(u)) \ne Y(u)$. Therefore there
is no $x_{\gamma'} \in X(U)$ such that $X(u) = x_{\gamma'} \Rightarrow
f_{\gamma'}(\Gamma(u)) = Y(u)$.  So our hypothesis is wrong; there is
no function that $C$ infers.

Now consider the case where $|\alpha| > 2$. Label the two elements of
$\beta$ as +1 and -1. Since $\alpha$ is a fine-graining of $\beta$,
and since $|\beta| = 2$, there are at least two distinct elements of
$\alpha$ that are contained in the same element of $\beta$, having
label $b$.  Choose one of those elements of $\alpha$, $a$, and let
$a'$ be one of the other elements of $\alpha$ that are contained in
that element of $\beta$ with label $b$.

Form the union of $a$ with all elements of $\alpha$ that are contained
in the element of $\beta$ with label $-b$. That union is a proper
subset of all the elements of $\alpha$. Therefore it picks out a
proper subset of $U$, $W$. (Note that $W$ has non-empty overlap with
both both partition elements of $\beta$.)  So choose $\Gamma$ to be
binary-valued, with values given by $\Gamma(u) = b$ iff $u \in
W$. Then for $X(u) = a$, $\Gamma(u) = b = Y(u)$. On the other hand,
for $X(u) = a'$, $\Gamma(u) = -b = -Y(u)$. So for both probes $f$ of
$\Gamma$, there is a value $x$ such that $X = x \Rightarrow Y =
f(\Gamma)$. {\bf{QED.}}

$ $

\noindent {\bf{Proof of Thm. 1:}} Let $C_1$ and  
$C_2$ be the two devices.  Since $Y$ for any inference device is
surjective, $Y_2(U) = {\mathbb{B}}$, and therefore there are two
probes of $Y_2(U)$.  Since by hypothesis $C_1$ weakly infers $C_2$,
using the identity probe $f(y
\in {\mathbb{B}}) = y$ establishes that $\exists \; x_1$ s.t. $X_1(u)
= x_1 \Rightarrow Y_1(u = Y_2$. Similarly, since $C_2$ weakly
infers $C_1$, using the negation probe $f(y) = -y$ establishes
that $\exists \; x_2$ s.t. $X_2(u) = x_2 \Rightarrow Y_2(u) =
-Y_1(u)$. Finally, by the hypothesis of setup distinguishability,
$\exists \; u^* \in U$ s.t. $X_1(u^*) = x_1, X_2(u^*) =
x_2$. Combining, we get the contradiction $Y_1(u^*) = Y_2(u^*) =
-Y_1(u^*)$. {\bf{QED.}}

$ $

\noindent {\bf{Proof of Thm. 2:}} To establish (i), let $f$ be any probe of
$\Gamma(U)$. $C_2 > \Gamma \Rightarrow \exists \; x_2$ such that
$X_2(u) = x_2 \Rightarrow Y_2(u) = f(\Gamma(u))$. In turn, $C_1
\gg C_2 \Rightarrow \exists \; x_1$ such that $X_1 = x_1
\Rightarrow Y_1 = Y_2, X_2 = x_2$ (by choosing the identity
probe of $Y_2(U)$). Combining, $X_1 = x_1
\Rightarrow Y_1(\Gamma)$. So $C_1 > \Gamma$, as claimed in (i).

To establish (ii), let $f$ be any probe of $Y_3(U)$, and $x_2$ any
member of $X_3(U)$. $C_2 \gg C_3 \Rightarrow \exists \; x_2 \in
X_2(U)$ such that $X_2(u) = x_2 \Rightarrow X_3(u) = x_3, Y_2(u) =
f(Y_3(u))$. $C_1
\gg C_2$ then implies that $\exists \; x_1$ such that
$X_1(u) = x_1 \Rightarrow X_2(u) = x_2, Y_1(u) = Y_2(u)$ (by choosing
the identity probe of $Y_2(U)$). Combining, $X_1(u) = x_1 \Rightarrow
X_3(u) = x_3, Y_1(u) = f(Y_3(u))$, as desired.  {\bf{QED.}}

$ $

\noindent {\bf{Proof of Prop. 2:}} To establish the first claim, simply take $Y_2$ to
be the function $\Gamma$ in Prop. 1(ii).

To establish the second claim, focus attention on any $x_1 \in
X_1(U)$, and define $W \equiv X_1^{-1}(x_1)$. Choose $X_2$ so that
$X_2(u)$ take on a separate unique value for each $u \in W$, i.e.,
$\forall w \in , u \in U$ : $w \ne u$, $X_2(w) \ne X_2(u)$. 

First consider the case where $Y_1(W)$ has a single element, i.e.,
$Y_1(u)$ is the same bit across all $X_1^{-1}(x_1)$. Without loss of
generality take that bit to be 1.  Choose $Y_2(u) = 1$ for some $w'
\in W$, and $Y_2(u) = -1$ for all other $w \in W$. Then choose $x_2$
so that $X_2(u) = x_2 \Rightarrow u = w'$. Therefore $X_2(u) = x_2
\Rightarrow X_1(u) = x_1, Y_2(u) = 1$. So for the probe $f$ of
$Y_1(U)$ that equals $Y_1$, $X_2(u) = x_2
\Rightarrow Y_2(u) = f(Y_1(u))$. On the other hand, by hypothesis
$\exists \; w'' \in W$ that differs from $w'$, and $\exists \; x_2'
\in X_2(U)$ such that $X_2(u) = x'_2 \Rightarrow u = w''$. Moreover,
$Y_2(w'') = -1$, by construction of $Y_2$. So consider the probe $f'$
of $Y_1(U)$ that equals $-Y_1$. For all $u \in W$, $f'(Y_1(u)) =
-1$. In particular, this is the case for $u = w''$. Combining, $X_2(u)
= x_2' \Rightarrow X_1(u) = x_1, Y_2(u) = f'(Y_1(u))$. Since $f$ and
$f'$ are the only probes of $Y_1(U)$, there are no more cases to
consider for the situation where $Y_1(W)$ is a singleton.

If $Y_1(W)$ is not a singleton, since $W$ contains at least three
elements, there is a proper subset of $W$, $W'$, on which $Y_1$ takes
both values. So by Prop. 1(i) there is a device $C$ over $W$ that
infers the restriction of $Y_1$ to domain $W$. Define $(X_2, Y_2)$ to
be the same as that $C$ for all $u \in W$, with all members of
$X_2(W)$ given values that are not found in $X_2(U - W)$. Since
$X_1(w) = x_1$ for all $w \in W$, this means that $\forall \; f
\in  \pi(Y_1)$, $\exists \; x_2$ such that $X_2(u) = x_2
\Rightarrow X_1(u) = x_1, Y_2(u) = f(Y_1(u))$.

Combining, since $Y_1(X_1^{-1}(x_1))$ either is or is not a singleton
for each $x_1 \in X_1(U)$, we can build a ``partial'' device $C_2$ that
strongly infers $C_1$ for each region $X_1^{-1}(x_1)$.  Furthermore,
those regions form a partition of $U$. So by appropriately ``stitching
together'' the partial $C_2$'s built for each $x_1 \in X_1(U)$, we
build an aggregate device $C_2$ that strongly infers $C_1$ over all
$U$, as claimed. {\bf{QED.}}

$ $

\noindent {\bf{Proof of Thm. 3:}} Let $C_1$ and  
$C_2$ be two devices and hypothesize that they can strongly infer each
other. Since $C_1$ can strongly infer $C_2$, it can force $X_2$ to
have any desired value and simultaneously correctly infer the value of
$Y_2$ under the identity probe. In other words, there is a function
$\xi^1_I : X_2(U)
\rightarrow X_1(U)$ such that for all $x_2$, $X_1
= \xi^1_I(x_2) \Rightarrow X_2 = x_2$ and $Y_1 = Y_2$. Let
${\hat{x}}_1$ be any element of $\xi^1_I(X_2(U))$.  

Similarly, by hypothesis $C_2$ can force $X_1$ to have any desired
value and simultaneously correctly infer the value of $Y_1$ under the
negation probe. In other words, there is a function $\xi^2_{-I} :
X_1(U) \rightarrow X_2(U)$ such that for all $x_1$, $X_2 =
\xi^2_{-I}(x_1) \Rightarrow X_1 = x_1$ and $Y_1 = -Y_2$.

Define ${\hat{x}}_2 \equiv \xi^2_{-I}({\hat{x}}_1)$.
Then $X_1(u) = \xi^1_I({\hat{x}}_2) \Rightarrow X_2(u) = {\hat{x}}_2 =
\xi^2_{-I}({\hat{x}}_1)$ and $Y_1(u) = Y_2(u)$. The first of those two
conclusions in turn means that $Y_1(u) = -Y_2(u)$. Combining, we see
that $X_1(u) = \xi^1_I({\hat{x}}_2)
\Rightarrow Y_2(u) = Y_1(u) = -Y_2(u)$, which is
impossible. {\bf{QED}}

$ $

\noindent {\bf{Proof of Thm. 4:}} Since $C_2 > \Gamma$, $\forall \; f \in
\pi(\Gamma)$, $\exists \; x_2$ such that $X_2 = x_2
\Rightarrow Y_2 = f(\Gamma)$.  Therefore the set argmin$_{x_2 : X_2
= x_2 \Rightarrow Y_2 = f(\Gamma)} [{\mathscr{L}}(x_2)]$ is
non-empty. Accordingly, $\forall f \in \pi(\Gamma)$, we can define an
associated value $x_2^f \in X_2(U)$ as some particular element of
argmin$_{x_2 : X_2 = x_2 \Rightarrow Y_2 = f(\Gamma)}
[{\mathscr{L}}(x_2)]$.

Now since $C_1 \gg C_2$, $\forall x_2$, $\exists \; x_1$ such that
$X_1 = x_1 \Rightarrow X_2 = x_2, Y_1 = Y_2$. In particular, $\forall
f \in \pi(\Gamma)$, $\exists \; x_1  : X_1 = x_1 \Rightarrow
X_2 = x^f_2, Y_1 = Y_2$. So by definition of $x_2^f$, $\forall f \in
\pi(\Gamma)$, $\exists \; x_1 : X_1 = x_1 \Rightarrow X_2 = x^f_2, Y_1 = f(\Gamma)$.

Combining, $\forall f \in \pi(\Gamma)$,
\begin{eqnarray*}
{\mbox{min}}_{x_1 : X_1 = x_1
\Rightarrow Y_1 = f(\Gamma)} [{\mathscr{L}}(x_1)] \;\;\; &\le& \;\;\;  {\mbox{min}}_{x_1 : X_1 = x_1
\Rightarrow X_2 = x^f_2, Y_1 = Y_2} [{\mathscr{L}}(x_1)]. 
\end{eqnarray*}
Accordingly,
\begin{eqnarray*}
{\mathscr{C}}(\Gamma \mid C_1) - {\mathscr{C}}(\Gamma \mid C_2) \;\;\; &\le& \;\;\;
\sum_{f \in \pi(\Gamma)}  {\mbox{min}}_{x_1 : X_1 = x_1
\Rightarrow X_2 = x^f_2, Y_1 = Y_2} [{\mathscr{L}}(x_1) -
{\mathscr{L}}(x_2^f)] \\
&\le& \;\;\; 
\sum_{f \in \pi(\Gamma)}  {\mbox{max}}_{x_2} \left[ {\mbox{min}}_{x_1 : X_1 = x_1
\Rightarrow X_2 = x_2, Y_1 = Y_2} [{\mathscr{L}}(x_1) -
{\mathscr{L}}(x_2)] \right] \\
&=& \;\;\;
|\pi(\Gamma)| \; {\mbox{max}}_{x_2} \left[ {\mbox{min}}_{x_1 : X_1 = x_1
\Rightarrow X_2 = x_2, Y_1 = Y_2} [{\mathscr{L}}(x_1) -
{\mathscr{L}}(x_2)] \right]
\end{eqnarray*}
Using the equality $|\pi(\Gamma)| = |\Gamma(U)|$ completes the proof.
{\bf{QED.}}

$ $

\noindent {\bf{Proof of Thm. 5:}} By hypothesis, for any $x'_2 \in X_2(U)$,
$\exists \; x_1$ such that $X_1 = x_1 \Rightarrow X_2 = x'_2$. This is
true for any such $x'_2$. Write the function mapping any such $x'_2$
to the associated $x_1$ as $\xi_1$. Similarly, there is a function
$\xi_2$ that maps any $x_1 \in X_1(U)$ to an $x_2 \in X_2(U)$ such
that $X_2 = \xi_2(x_1) \Rightarrow X_1 = x_1$. Using the axiom of
choice, this provides us with a single-valued mapping from $X_1(U)$
into $X_2(U)$ and vice-versa.

Since having $X_2(u) = \xi_2(x_1)$ forces $X_1(u) = x_1$, the set of
$u \in U$ such that $X_2(u) = \xi_2(x_1)$ must be a subset of those $u
\in U$ such that $X_1(u) = x_1$, i.e., $\forall \; x_1$, 
$X_2^{-1}[\xi_2(x_1)] \subseteq X_1^{-1}(x_1)$. Similarly, $\forall \;
x_2$, $X_1^{-1}[\xi_1(x_2)] \subseteq X_2^{-1}(x_2)$. This second
equality means in particular that $X_1^{-1}[\xi_1[\xi_2(x_1))]
\subseteq X_2^{-1}(\xi_2(x_1))$. Combining, $X_1^{-1}[\xi_1[\xi_2(x_1))]
\subseteq X_1^{-1}(x_1)$.

However $\forall \; x_1$, $\xi_1(\xi_2(x_1))$ is
non-empty. Since $X_1$ is single-valued, this means that $\forall \;
x_1$, $\xi_1(\xi_2(x_1)) = x_1$. Combining, we see that
$\forall \; x_1$, $X_1^{-1}(x_1) \subseteq
X_2^{-1}[\xi_2(x_1)]$, and therefore $X_2^{-1}[\xi_2(x_1)] =
X_1^{-1}(x_1)$. This in turn means that the set $X_2[X_1^{-1}(x_1)]$
equals the singleton $\xi_2(x_1)$ for any $x_1 \in
X_1(U)$. Accordingly $\forall \; u \in X_1^{-1}(x_1)$, $X_2(u) =
\xi_2(x_1) =  \xi_2(X_1(u))$. In addition, every $u \in U$ obeys $ u
\in X_1^{-1}(x_1)$ for $some$ $x_1$. Therefore we conclude
that for all $u \in U$, $\xi_2(X_1(u)) = X_2(u)$.  

This establishes that the partition induced by $X_1$ is a fine-graining
of the partition induced by $X_2$. Similar reasoning establishes that
the partition induced by $X_2$ is a fine-graining of the partition
induced by $X_1$. This means that the two partitions must be
identical. {\bf{QED.}}

$ $

\noindent {\bf{Proof of Coroll. 3:}} By Thm. 5, we can relabel the image values
of the two devices' setup functions to express them as $C_1 = (X,
Y_1)$ and $C_2 = (X, Y_2)$.

To prove (i), note that $C_1 > C_2$ means $\exists \; x \in X(U)$ such
that $X = x \Rightarrow Y_1 = Y_2$ and $\exists \; x' \in X(U)$ such
that $X = x' \Rightarrow Y_1 = -Y_2$. But those two properties in turn
mean that $C_2 > C_1$. A similar argument establishes that $C_2 > C_1
\Rightarrow C_1 > C_2$.

To prove (ii), note that $C_1 \gg C_2$ means that $\forall x \in X(u),
f \in \pi(Y_2)$, $\exists \; x'$ such that $X = x' \Rightarrow X = x,
Y_1 = f(Y_2)$. In particular, $\forall x \in X(u)$, $\exists \; x'$
such that $X = x' \Rightarrow X = x, Y_1 = Y_2$, and $\exists \; x''$
such that $X = x'' \Rightarrow X = x, Y_1 = -Y_2$. The only way both
conditions can hold is if $x' = x''$. But that means it is impossible
to have both $Y_1 = Y_2$ and $Y_1 = -Y_2$. 

To prove (iii), hypothesize that $C_1$ control $X$. This means in
particular that $\forall x \in X(U)$, $\exists \; x' \in X(U)$ such
that $X = x' \Rightarrow Y_1 = \delta_{X, x} = 1$ (choose $b = 1$ and
have $f$ be the probe that equals 1 iff its argument equals $x$). To
have $\delta_{X, x} = 1$ means $X = x$, which in turn means $x' =
x$. So $X = x \Rightarrow Y_1 = 1$. This is true for all $x \in X(U)$,
so $Y_1(u) = 1 \; \forall u \in U$. However by definition, the range
of $Y_1$ must be $\mathbb{B}$. Therefore the hypothesis is wrong. The
same argument shows that $C_2$ cannot control $X$. {\bf{QED.}}

$ $

\noindent {\bf{Proof of Thm. 6:}} To prove (i), let $f$ be any probe of
$\Gamma$. Intelligibility means $f \in Q_1(U)$.  Since $D_1$
semi-controls its question function, $\exists x_1 : X_1 = x_1
\Rightarrow Q_1 = f$. Infallibility then implies that for any $u$ such
that $X_1(u) = x_1$, $Y_1(u) = [Q_1(u)](u) = f(u)$. This proves (i).

Next, let $f$ be any probe of $Y_2$, and $x_2$ any element of
$X_2(U)$. Intelligibility means $f \in Q_1(U)$.  Since $D_1$
semi-controls $(Q_1, X_2)$ and $(Q_1, X_2)$ is surjective, $\exists
x_1$ such that $X_1 = x_1 \Rightarrow Q_1 = f, X_2 =
x_2$. Infallibility then implies that for any $u$ such that $X_1(u) =
x_1$, $Y_1(u) = [Q_1(u)](u) = f(u)$. This proves (ii). {\bf{QED.}}

$ $

\noindent {\bf{Proof of Thm. 7:}} The cardinality of $\pi(P)$ is the
cardinality of $P(U)$, $|P(U)|$. Let $f_1$ and $f_2$ be two separate
such probes, so that $f_1 : P(U) \rightarrow {\mathbb{B}}$ differs
from $f_2 : P(U) \rightarrow {\mathbb{B}}$. Then as functions over
$U$, $f_1(P)$ and $f_2(P)$ differ. Therefore by hypothesis they
correspond to two distinct $q$'s in $Q'(U)$. So $|Q'(U)| \ge
|P(U)|$. In turn, $|Q(U)| = |R(P(U))| \le |P(U)|$. So $|Q'(U)| \ge
|Q(U)|$. Similar reasoning establishes that $|Q(U)| \ge |Q'(U)|$. So
$|Q(U)| = |Q'(U)|$. Therefore $|Q(U)| = |P(U)|$ and $|Q'(U)| =
|P'(U)|$. This proves (i).

Now since $P'$ is intelligible to $D$, every $f \in \pi(P')$ is an
element of $Q(U)$. Therefore for $|Q(U)|$ finite, (i)'s conclusion
that $|Q(U)| = |P'(U)|$ means that there is no $q \in Q(U)$ that is
not an element of $\pi(P')$. In other words, $Q = \pi(P')$. Next,
(i)'s conclusion that $|P'(U)| = |R'(P'(U))|$ establishes that the
partition induced by $P'$ is identical to the partition induced by
$R'(P')$. So $\pi(P') = \pi(Q')$. Similar reasoning establishes that
$Q' = \pi(P) = \pi(Q)$. This establishes (ii). {\bf{QED.}}

$ $

\noindent {\bf{Proof of Coroll. 4:}} Choose $P = (Y, Q)$ and $R : (Y, Q)(u)
\rightarrow Q(u)$. (So $R$ is a projection map.) Since $(Y, Q)$
is surjective, $|P(U)| = |(Y, Q)(U)| = 2|Q(U)|$. By Thm. 7, this is
impossible if the two self-aware devices are intelligible to each
another.  {\bf{QED.}}

$ $

\noindent {\bf{Proof of Prop. 3:}} The validity of the claim in (i) is independent
of the question function of the devices, so they can be set
arbitrarily. Choose $X_1(U) = X_2(U) = X_3(U) = \{0, 1\}$. Then
choose the reduced form of the setup and conclusion functions of the
devices in the reality to be the following four tuples: $([0, 0], [0,
0], [0, 0])$; $([0, 0], [[1, 0], [1, 1])$; $([1, 1], [0, 0], [1, 0])$;
$([1, 1], [1, 0], [0, 1])$. It is straightforward to verify that each
pair of devices is distinguishable and that $C_1 > C_2 > C_3 > C_1$.

To prove (ii), note that under hypothesis, $C_1 > C_2 \Rightarrow
\exists \; x_1 : X_1 = x_1 \Rightarrow Y_1 = Y_2$, $C_2 > C_3
\Rightarrow \exists \; x_2 : X_2 = x_2 \Rightarrow Y_2 = Y_3$,
$\ldots, C_{n-1} > C_n \Rightarrow
\exists \; x_{n-1} : X_{n-1} = x_{n-1} \Rightarrow Y_{n-1} = Y_n$,
$C_{n} > C_1 \Rightarrow 
\exists \; x_{n} : X_{n} = x_{n} \Rightarrow Y_{n} = -Y_1$ . Mutual
distinguishability means that there is a tuple in the reduced form
of the reality having that set of $x_i$ values. However that would
mean that the tuple has $y_1 = -y_1$. So our hypothesis is wrong.

To prove (iii), simply combine Thm. 3 and Thm. 2.  {\bf{QED.}}

$ $

\noindent {\bf{Proof of Prop. 4:}} Since $D$ is acyclic and finite, it contains at
least one root node. Label one such node as $C_1$. Hypothesize that
there is some other root node in the graph.

Given any $D' \subseteq D$, define $S(D')$ as the union of $D'$ with
the set of all nodes in $D$ that are successors of a node in
$D'$. Similarly, define $P(D')$ as the union of $D'$ with the set of
all nodes in $D$ that are predecessors of a node in $D'$.  $S(\{C_1\})
\subset D$ since by hypothesis there is more than one root node. Since
$D$ is weakly connected, this means that $S(\{C_1\}) \subset
P[S(\{C_1\})]$. Since $D$ is acyclic and finite, this means that there
is a node $C_j \in S(\{C_1\})$ who has a root node predecessor $C_k$
where $C_k \not \in S(\{C_1\})$.

So $C_j$ is a successor of two separate root nodes, $C_k$ and $C_1$. By
transitivity of strong inference, this means that $C_1 \gg C_j$ and $C_k
\gg C_j$. By the hypothesis of the proposition, since $C_k \ne C_1$,
those two devices are distinguishable. This means it is possible for
$C_1$ to force $X_j$ to have one value while at the same time $C_k$
forces $X_j$ to have a different value. This is a contradiction.
{\bf{QED.}}

$ $

\noindent {\bf{Proof of Prop. 5:}} The proof of (i) is by
example. Consider the following set of five quadruples:
\begin{eqnarray*}
V \;&\equiv&\; \{(-1, -1, -1, -1); (-1, -1, 1, -1); (1, -1, -1, 1);
(1, 1, 1, -1), (-1, 1, 1, 1)\}
\end{eqnarray*}
By Lemma 1, $V$ is the reduced form of a reality consisting of two
devices $C_1$ and $C_2$, where we identify any quadruple in $V$ as the
value $(x_1, y_1, x_2, y_2)$, so that $X_1(U) = X_2(U) =
{\mathbb{B}}$. By inspection, $C_1 > C_2$ (e.g., $X_1 = 1 \Rightarrow
Y_1 = -Y_2$). Similarly, by inspection $C_1$ and $C_2$ are
distinguishable, and copies of each other. This completes the proof of
(i).

To prove the first part of (ii), first note that $C_1 \gg C_2$
requires that for all $x_2$, there is (an $x_1$ that forces $X_2 =
x_2$ and $Y_1 = Y_2$), and (an $x_1$ that forces $X_2 = x_2$ and $Y_1
= -Y_2$). In other words, there is a single-valued map $\xi : X_2(U)
\rightarrow X_1(U)$ such that the quadruple $(X_1 = \xi(x_2), Y_1 =
y_1, X_2 = x_2, Y_2 = y_1)$ occurs for some $y_1$ in some tuple in the
reduced form of the reality while $(X_1 = \xi(x_2), Y_1 = y_1, X_2 =
x'_2, Y_2 = y_2)$ does not occur for any $y_2$ if $x'_2 \ne x_2$, and
also does not occur for $y_2 = -y_1$ if $x'_2 = x_2$. Similarly, there
is a single-valued map $\xi' : X_2(U) \rightarrow X_1(U)$ such that
the quadruple $(X_1 = \xi(x_2), Y_1 = y_1, X_2 = x_2, Y_2 = -y_1)$
occurs for some $y_1$ in some tuple in the reduced form of the reality
while $(X_1 = \xi(x_2), Y_1 = y_1, X_2 = x'_2, Y_2 = y_2)$ does not
occur for any $y_2$ if $x'_2 \ne x_2$, and also does not occur for
$y_2 = y_1$ if $x'_2 = x_2$. By construction, both $\xi$ and $\xi'$
are invertible. Furthermore, for all $x_2$, $\xi(x_2) \ne \xi'(x_2)$.
So $|X_1(U)| \ge 2|X_2(U)|$. On the other hand, $|X_1(U)| = |X_2(U)|$
because $C_1$ and $C_2$ are copies of each other. Therefore they must
have infinite setup functions.
  
The existence proof for (ii) is by example. Define a set of quadruples
\begin{eqnarray*}
T &\equiv& \{(1, -1, 1, -1); (2, 1, 1, -1); (3, -1, 2, 1); (4, 1, 2, 1);
(5, -1, 3, -1), (6, 1, 3, -1), \ldots\} \nonumber \\
&=& \{(i, \;1 - 2(i {\mbox{ mod }} 2),\;
\lceil(i/2), \; 1 - 2(\lceil(i/2) {\mbox{ mod }} 2)) \; : \: i \in
{\mathbb{N}}\}
\end{eqnarray*}
Next, fix any set of spaces $\sigma$, where the spaces $\{y_1\} =
\{y_2\} \equiv {\mathbb{B}}$ and $\{x_1\} = \{x_2\} \equiv \mathbb{N}$
all occur in $\sigma$.  Let $S$ be a subset of the Cartesian product
of the spaces in $\sigma$.  Say that for every $t
\in T$, $(x_1, y_1, x_2, y_2) = t$ for exactly one element of $S$, and
no element of $S$ contains a quadruple $(x_1, y_1, x_2, y_2) \not
\in T$.  (So there is a bijection between $S$ and $T$, given by
projecting any element of $S$ onto its four components corresponding
to the spaces $\{x_1\}, \{x_2\}, \{y_1\}$ and $\{y_2\}$.)

By Lemma 1, $S$ is the reduced form of a reality, where we can define
$X_1(U) \equiv \{x_1\}, Y_1(U) \equiv \{y_1\}, X_2(U) \equiv \{x_2\},
Y_2(U) \equiv \{y_2\}$.  Accordingly group $(X_1, Y_1)$ into a device
$C_1$ and $(X_2, Y_2)$ into a device $C_2$.  By inspection, the
relation in $T$ between pairs $x_1$ and $y_1$ is identical to the
relation in $T$ between pairs $x_2$ and $y_2$. (Those relations are
the pairs $\{(1, -1); (2, 1); (3, -1),
\ldots\}$.) So the devices $C_1$ and $C_2$ in the reality are copies
of each other.

Next, note that $\forall  x_2 \in {\mathbb{N}}, y_1 \in
{\mathbb{B}}$, $(2x_2 +
\frac{(y_1 - 1)}{2}, y_1, x_2, 1 - 2 (x_2 {\mbox{ mod }} 2))$ occurs
(once) in $T$. Accordingly, $X_1 = 2x_2 + \frac{(y_1 - 1)}{2}
\Rightarrow X_2 = x_2$. Also, for any fixed $x_2$, choosing either
$X_1 = 2x_2$ or $X_1 = 2x_2 - 1$ forces $y_1$ to be either $1$ or
$-1$, respectively. Therefore, given that $x_2$ is fixed, it also
forces either $y_1 = 1 - 2 (x_2 {\mbox{ mod }} 2)$ or $-y_1 = 1 - 2
(x_2 {\mbox{ mod }} 2)$.  (For example, $X_1 = 5$ forces $X_2 = 3$ and
$Y_1 = Y_2$, while $X_1 = 6$ forces $X_2 = 3$ and $Y_1 = -Y_2$.) So
the choice of $X_1$ forces either $Y_1 = Y_2$ or $Y_1 =
-Y_2$. Therefore $C_1 \gg C_2$. {\bf{QED}}.

$ $

\noindent {\bf{Proof of Prop. 6:}} Plugging in, the product of the
two inference accuracies is
\begin{eqnarray*}
\left(\frac{\sum_{f_1 \in \pi(Y_2)}{\mbox{max}}_{x_1} [{\mathbb{E}}_P(Y_1 f_1(Y_2)
\mid x_1)]}{2}\right) \left(
\frac{\sum_{f_2 \in \pi(Y_1)}{\mbox{max}}_{x_2} [{\mathbb{E}}_P(Y_2 f_2(Y_1) \mid x_2)]}{2}\right).
\end{eqnarray*}
Define $g \equiv Y_1 Y_2$. Then we can rewrite our product as
\begin{eqnarray*}
\left(\frac{{\mbox{max}}_{x_1} [{\mathbb{E}}_P(g \mid x_1)]}{2} \;+\;
  \frac{{\mbox{max}}_{x_1} [{\mathbb{E}}_P(-g \mid x_1)]}{2}\right) \left(
\frac{{\mbox{max}}_{x_2} [{\mathbb{E}}_P(g \mid x_2)]}{2} \;+\;
  \frac{{\mbox{max}}_{x_2} [{\mathbb{E}}_P(-g \mid x_2)]}{2}\right).
\end{eqnarray*}
For $|X_1(U)| = |X_2(U)| = 2$, we can rewrite this as
\begin{eqnarray*}
\left(\frac{|{\mathbb{E}}_P(g \mid X_1 = 1) \;-\; {\mathbb{E}}_P(g \mid
  X_1 = -1)|}{2}\right) \left(
\frac{|{\mathbb{E}}_P(g \mid X_2 = 1) \;-\; {\mathbb{E}}_P(g \mid X_2
  = -1)|}{2}\right).
\end{eqnarray*}

Next, since the distinguishability is 1.0, $X_1$ and $X_2$ are
statistically independent under $P$. Therefore we can write $P(g, x_1,
x_2) = P(g \mid x_1, x_2)P(x_1)P(x_2)$.  So for example, $P(g \mid
x_1) = \sum_{x_2}P(g \mid x_1, x_2) P(x_2)$, and 
\begin{eqnarray*}
{\mathbb{E}}_P(g \mid x_1) &=& \sum_{x_2}[P(g = 1 \mid x_1, x_2) - P(g = -1 \mid x_1, x_2)]
P(x_2) \nonumber \\
&=& 2[\sum_{x_2}P(g = 1 \mid x_1, x_2)P(x_2)] - 1.
\end{eqnarray*}

Now define $z_1 \equiv P(g = 1
\mid x_1 = -1, x_2 = -1), z_2 \equiv P(g = 1 \mid x_1 = -1, x_2 = 1), z_3
\equiv P(g = 1 \mid x_1 = 1, x_2 = -1), z_4 \equiv P(g = 1 \mid x_1 = 1,
x_2 = 1)$. Note that the 4-tuple $(z_1, z_2, z_3, z_4) \in H$ so long
as none of its components equals 0.
Plugging in,
\begin{eqnarray*}
{\mathbb{E}}_P(g \mid X_1 = -1) &=& 2[z_1 \beta + z_2(1 - \beta)] - 1,
\nonumber \\
{\mathbb{E}}_P(g \mid X_1 = 1) &=& 2[z_3 \beta + z_4(1 - \beta)] - 1,
\nonumber \\
{\mathbb{E}}_P(g \mid X_2 = -1) &=& 2[z_1 \alpha + z_3(1 - \alpha)] - 1,
\nonumber \\
{\mathbb{E}}_P(g \mid X_2 = 1) &=& 2[z_2 \alpha + z_4(1 - \alpha)] -
1.
\end{eqnarray*}
So the product of inference accuracies is
\begin{eqnarray*}
| [\beta(k({\vec{z}})) + m({\vec{z}})][\alpha(k({\vec{z}}) +
n({\vec{z}})]| \;&=& \;
| \alpha \beta [k({\vec{z}})]^2 + \alpha k({\vec{z}})m({\vec{z}}) +
\beta k({\vec{z}})n({\vec{z}}) + m({\vec{z}})n({\vec{z}})| .
\end{eqnarray*}

This establishes the first part of the proposition. Note that
depending on the structure of the mapping from $(X_1, X_2) \rightarrow
(Y_1, Y_2)$, if we require that both $Y_i$ be stochastically
surjective, there may be constraints on which quadruples $z \in H$ are
allowed. Such restrictions would make our bound be loose.

When $\alpha = \beta = 1/2$, the product of inference accuracies
reduces to
\begin{eqnarray*}
|\frac{z_1^2 - z_2^2 - z_3^2 + dz_4^2}{4} + \frac{z_2 z_3 -
z_1z_4}{2}| \;&=&\; |\frac{(z_1 - z_4)^2 - (z_2 - z_3)^2}{4}|
\end{eqnarray*}
This establishes the second claim. The final claim is established by
maximizing this expression over $H$. 
{\bf{QED.}}

$ $

\noindent {\bf{Proof of Prop. 7:}} Given any $C_1 = (X_1, Y_1)$, the proposition
is proven if we can construct an associated $D_2$ that $C_1$ does not
correct. To do that, choose $Y_2 = Y_1$, and have $Q_2(U)$ consist of
two elements, $q_1 = Y_1$, and $q_2 = -Y_1$.  Define $Q_2$'s
dependence on $u \in U$ by requiring that $Y_1 = -1 \Leftrightarrow
Q_2 = {\underline{q_1}}$ (i.e., $\forall u \in U$ such that $Y_1(u) =
-1$, $Q_2(u) = q_1 = Y_1$), and by requiring that $Y_1 = 1
\Leftrightarrow Q_2 = {\underline{q_2}}$. (Since $Y_1$ is surjective
onto ${\mathbb{B}}$, this defines $Q_2$'s dependence on all of $U$,
and guarantees that $|Q_2(U)| \ge 2$, as required.)

Plugging in, ${\overline{Q_2}} = {\underline{-1}}$. Now the square of
both 1 and -1 equals 1. Since $Y_1 = Y_2$, this means that $Y_1 Y_2 =
{\underline{1}}$. Combining, ${\overline{Q_2}} = -Y_2 Y_1$. Therefore
$Y_2 {\overline{Q_2}} = -Y_1$. Therefore it is impossible that $Y_1 =
Y_2 {\overline{Q_2}}$, i.e., there is no $x_1$ that implies this
equality.  {\bf{QED.}}

$ $

$ $

\noindent {\bf{APPENDIX B: The lack of restrictions in the definition
of weak inference}} 

$ $

Note that there is additional structure in Ex. 1 that is missing in
Def. 3. Most obviously, no analog of $\zeta$ appears in Def. 3. In
addition, Def. 3 does not require that there be a component of $X$
and/or $Y$ that can be interpreted as a question-valued function like
$Q$. Moreover, even if it is the case that $X = \chi \otimes Q$,
Def. 3 allows the value imposed on $\chi$ to vary depending on what
probe one is considering, in contrast to the case in
Ex. 1. Alternatively, it may be that the question $Q(u)$ does not
equal the associated probe $f_K$ that is being answered, but so long
as $Y(u) = f_K(\Gamma(u))$ whenever $\chi(u) \otimes Q(u)$ has a
certain value, the device ``gets credit'' for being able to answer
question $f_K$. In this, the definition of weak inference doesn't
fully impose the mathematical structure underpinning the concept of
semantic information. Phrased differently, the impossibility results
for weak inference hold even though weak inference only uses some of
the structure needed to define semantic information. (See
Sec.~\ref{sec:sad} for results that involve all of that structure.)

In addition, it may be that the scientist cannot read the apparatus'
output display accurately. In this case the scientist would give
incorrect answers as to what's on that display. However so long as
that inaccuracy was compensated, say by a mistake in the observation
apparatus, we would still say that the device infers $\Gamma$.  Any
such extra structure that is in Ex. 1 can be added to the definition
of weak inference in Def. 3 if desired, and the impossibility results
presented here for weak inference will still obtain. (See
Sec.~\ref{sec:sad} for a formalization of inference that contains
additional structure much like that in Ex. 1.)

The other examples in Sec.~\ref{sec:examples} can be cast as instances
of weak inference in similar fashions.  In particular, all of them
have additional structure beyond that required in Def. 3.

It is worth elaborating further this point of just how unrestrictive
Def. 3 is. One might argue that to apply to things like computers
being used for prediction, a definition of inference should involve
additional formal structure like time-ordering, or stipulations about
the Chomsky hierarchy power of the device, or stipulations about
physical limits restricting the device's operation like the speed of
light, quantum mechanical uncertainties, etc..  More abstractly, one
might argue that for a conclusion of a device to be physically
meaningful, it should be possible to ``act'' upon that conclusion, and
then test through the universe's response to that action whether the
conclusion is correct. None of this is required.

Note also that Def. 3 doesn't require that the device be used to infer
some aspect of world ``outside'' of the device. For example, no
restrictions are imposed concerning the physical coupling (or lack
thereof) at any particular instant of time between the device and what
the device infers. The device and what it is inferring can be anything
from tightly coupled with each other to completely isolated from each
other, at any moment.

As an extreme version of the first end of that spectrum, one can even
have the device and what it is inferring be ``the same system''. For
example, this is the case if $X$ and/or $Y$ depend on every degree of
freedom in the universe at some moment in time (in some associated
reference frame). In such a situation, the entire universe is the
inference device, and it is being used to infer something concerning
itself.

As another example of the generality of the definition, note that time
does not appear in Def. 3. Ultimately, this is the basis for the fact
that the definition of inference applies to both prediction and
recollection, aka ``retrodiction''. This absence of time in Def. 3
also means that not only might the device be the entire universe, but
it might be the entire universe across all time.  In such a situation,
the device is not localized either spatially or physically; the setup
and/or conclusion of the device is jointly specified by all degrees of
freedom of the universe at all moments.

In addition, $X = x \Rightarrow Y = f({\Gamma})$ does not mean that
$Y(u)$ is the same for every $u \in X^{-1}(x)$. It simply means that
whatever values $Y(u)$ has as $u$ varies across $X^{-1}(x)$ are the
same as the values that $f(\Gamma(u))$ has. This weakness in the
definition of inference is necessary for it to accommodate observation
devices. (Recall that in such devices $X(u)$ is how the observation
device is set up, and the conclusion of the device depends on
characteristics of the external universe, to be types of inference
devices.)

Along the same lines, $C > \Gamma$ does not imply that there is
exactly one probe of $\Gamma$ for which the associated conclusion
value is 1. (This is true even though $\pi(\Gamma(U))$ is a full unary
representation of $\Gamma(U)$.) Formally, $C > \Gamma$ does not imply
that there is exactly one probe $f$ of $\Gamma$ such that $\exists \;
x : X = x \Rightarrow Y = f(\Gamma) = 1$. There may be more than one
such $f$, or even none.  So as embodied in weak inference, for $C$ to
predict (something concerning the future state of the universe as
encapsulated in the function) $\Gamma$ does not mean that for each
$\gamma \in \Gamma(U)$ there is some associated question $x$ that if
embodied in $X$ guarantees that $Y$ correctly says, ``yes, in this
universe $u$, $\gamma$ is the value that will occur; $\Gamma(u) =
\gamma$''. Weak inference only requires that for each $\gamma$ and
associated probe, $X$ can be set up so that the device's answer $Y(u)$
must be correct, not that it can be set up to be correct and answer in
the affirmative.

Similarly, $C > \Gamma$ does not imply that $C$ can
infer a ``coarse-grained'' version of $\Gamma$. It implies that $C$
can correctly answer, ``does $\Gamma(u)$ equal $\gamma_1$?''  for some
$\gamma_1 \in \Gamma(U)$, and that it can correctly answer ``does
$\Gamma(u)$ equal $\gamma_2$'' for some $\gamma_2 \in
\Gamma(U)$. However it does not imply that $C$ can correctly answer,
``does $\Gamma(u)$ equal either $\gamma_1$ or $\gamma_2$ or
both?''. In particular, for two functions over $U$, $\Gamma$ and
$\Gamma'$, $C > (\Gamma, \Gamma')$ does not imply $C > \Gamma$. 

As another example of how weak Def. 3 is, recall that $Y$ is to be
interpreted as including all that the device ``knows''. On the other
hand, it is $X$ that includes a specification of what inference task
the device is being asked to perform. So in the definition of
inference, we don't even require that the device knows what inference
task it is being asked to perform. We just ask if it can be given such
a task and then come to the right conclusion, even if it doesn't know
what its conclusion ``means''.

There is no reason one could not introduce additional formal structure
in the definition of inference to embody some (or all) of these
attributes. For example, say we want to analyze the property of a
device $C$ both inferring some $\Gamma$ while also being capable of
correctly answering ``does $\Gamma(u)$ equal either $\gamma_1$ or
$\gamma_2$ or both?''. We could do this by strengthening the
definition of weak inference to also require that for any union of
probes of $\Gamma$, $\Phi$, there is an $x \in X(U)$ such that $X(u) =
x$ implies that $Y(u) = 1 \Leftrightarrow f(\Gamma(u)) = 1$ for some
$f \in \Phi$. (In general the $x \in X(U)$ that force the device to
infer such unions of multiple probes are different from the $x \in
X(U)$ that force the device to infer single probes.)  As another
example, say we want to have $C$ infer some $\Gamma$ while also
knowing how it is set up (so in particular it knows what probe of
$\Gamma$ it is inferring). We can accomplish this by requiring $C >
(\Gamma, X)$.

Whatever difficulties such additional structures might impose, they
are in addition to the impossibility results we derive below; the
results below apply no matter what such additional structures are
imposed.

In addition, in Def. 3 there are no restrictions on how, physically,
the function $\Gamma$ gets mapped to the setup value $x$. So there are
no stipulations, implicit or otherwise, about how $x$ is
interpreted. A mechanism for forcing $X(u)$ to have the desired value
for its inference will typically exist in any real device. In fact, in
general to infer different functions will require different such
mechanisms. So in the real world there is typically a way to replace
one such mechanism with another, depending on the function $\Gamma$
being inferred.

By leaving the mechanism out of the formal definition of inference,
all such complications are avoided.  In Def. 3, we simply say there
exists some appropriate $x \in X(U)$ for any $f(\Gamma)$, with nothing
mentioned about how to force the inference device (and therefore $u$)
to have what the device is supposed to compute, $f(\Gamma)$, reflected
in the value $x$.

Indeed, given any device $C$, we can define a new device
$C' \equiv (X', Y')$ where $X'(u)$ itself specifies the $f(\Gamma)$
that we wish to answer using the original device $(X, Y)$. So for
example, say $(X, Y)$ is a computer running a physical simulation
program whose initialized state is given by $X(u)$. Then $C'$ is that
computer modified by having a ``front end'' program that runs first to
figure out how to initialize the simulation to have the bit it
produces as a conclusion answer the question of interest. In this
case, trivially, there is no issue in mapping from $\Gamma$ to $x$;
that mapping is part of the setup function of our new device, $X'(.)$.

In particular, say that there is an ``external'' scientist who types
into the computer $C$ a specification of the system whose evolution is
to be simulated in the computer (i.e., forces $X(u)$ to have a value
that is interpreted as that specification). Then one can define $C'$
so that the scientist is embodied in $X'(.)$. In this definition, we
view the human scientist as ``part of'' the device (s)he is using.

In summary, and speaking very colloquially, one can view weak
inference as a necessary condition for saying that a device ``knows''
the actual value of a function of the state of the universe. Whatever
else such knowledge entails, it means that the device can, by whatever
means, correctly answer (with a yes or a no), ``Does the value of the
function of the state of the universe equal $z$?'' for any value $z$
in the codomain of the function.

Like with weak inference, there is no requirement that a device knows
how it has been set up for it to strongly infer another
device. Similarly, there is no requirement that it be able to strongly
infer the unions of probes, no requirements concerning its position in
the Chomsky hierarchy, etc. Despite being so pared-down, the
definition of strong inference is still sufficient to exhibit some
non-trivial behavior.

$ $

$ $

\noindent {\bf{APPENDIX C: Alternative definitions of weak inference}}

$ $

There are alternatives to Def. 3 that accommodate the case where
$|\Gamma(U)| > 2$ without employing multiple probes.  One such
alternative uses multiple devices in concert, each sharing the same
setup function, and each device's conclusion giving a different bit
concerning $\Gamma$'s value. As an example, say that $\Gamma$'s range
is $\mathbb{R}$. Then we could assign each device to a separate real
number, and require that for all $u$ one and only one device's
conclusion equals 1, namely the device corresponding to the value of
$\Gamma(u)$. 

To formalize this, say we have a set of devices \{$C_z : z \in
{\mathbb{R}}$\} and some function $\Gamma : U \rightarrow
{\mathbb{R}}$. In addition suppose there is some vector ${\vec{x}}$
with components $x_z$ running over all $z \in {\mathbb{R}}$ such that

$ $

\noindent {\bf{i)}} $\cap_{z \in {\mathbb{R}}} X^{-1}_z(x_z) \; \equiv
\; {\hat{U}}_\Gamma \; \ne\; \varnothing$.

$ $

\noindent {\bf{ii)}} $u \in {\hat{U}}_\Gamma \; \Rightarrow \; \forall
z \in {\mathbb{R}}$, $Y_z = 1$ iff $\Gamma(u) = z$.

$ $

\noindent {\bf{iii)}}  $\forall \gamma \in \Gamma(U)$, $\exists u \in
{\hat{U}}_\Gamma$ such that $Y_\gamma(u) = 1$.

$ $

\noindent Then we can jointly set up the set of devices so that their
joint conclusion gives $\Gamma(u)$, and we can do so without
precluding any element of $\Gamma(u)$. In this, the set of devices
``jointly infers'' $\Gamma$.

Alternatively, we could use a single device, where we modify the
definition of ``device'' to allow arbitrary cardinality of the range
of $Y$. With this modification, the conclusion function of the device
does not answer the question of what the value of a particular
function of $\Gamma(U)$ is. Rather it directly encodes the value of
$\Gamma(U)$.

It would appear that under such an alternative we do not need to have
the value of $X(u)$ specify the bit concerning $\Gamma(u)$ that we
want to infer, and do not need to consider multiple probes.  So for
example, it would appear that when the device is being used for
prediction, under this alternative $X(u)$ need only specify what is
known concerning the current state of the system whose future state is
being predicted, without specifying a particular bit concerning that
future state that we wish our device to predict. The conclusion $Y$
(or set of conclusions, as the case might be) would specify the
prediction in full.

Things are not so simple unfortunately. If we wish to allow the device
to infer functions $\Gamma$ with different ranges, then under this
alternative we have to allow different functions relating $Y(u)$ and
$\Gamma(u)$. This need is especially acute if we want to allow
$|\Gamma(U)|$ to vary.

Such functions should be surjective, to ensure that our device can
conclude every possible value of $\Gamma(U)$. (This surjectivity is
analogous to the requirement that we consider all probes in Def. 3.)
For any such function $\phi : Y(U)
\rightarrow \Gamma(U)$, we would interpret a particular value $Y(u)$
as saying ``$\Gamma(u) =
\phi(Y(u))$''. (This contrasts with the situation when $Y(U)
= {\mathbb{B}}$, where we interpret $Y(u) = +1/$$-1$ to mean ``yes/no'',
respectively, in response to the question of whether some associated
probe has the value +1.)

One immediate problem with this alternative definition of inference is
that it does not allow a device $(X, Y)$ to infer any function
$\Gamma(U)$ where $|\Gamma(U)| > |Y(U)|$. Such difficulties do not
hold for Def. 3. For example, if $X(U) = 3$, $X$ is a fine-graining of
$Y$ with two of its elements contained in $Y^{-1}(-1)$, and $\Gamma$
is a fine-graining of $X$, then $(X, Y) >
\Gamma$. (For every probe of $\Gamma(U)$, $x$ is chosen to be one of
the two elements that cause $Y(u) = -1$. The precise $x$ chosen for a
particular probe $f$ is the one that lies in $(f(\Gamma))^{-1}(-1)$.)

Other difficulties arise when we try to specify this alternative
definition in full. For example, one possible such definition is that
$C$ infers $\Gamma$ iff $\exists
\; x$ and function $\phi : Y(U) \rightarrow \Gamma(U)$ such that $X = x
\Rightarrow \phi(Y) = \Gamma$. Such a definition is unsatisfying in
that by not fixing $\phi$ ahead of time, it leaves unspecified how the
conclusion of the device is to be physically interpreted as an
encoding of $\Gamma(u)$. (This is in addition to the lack of a fixed
mapping from $\Gamma$ to $x$, a lack which also arises in Def. 3.)

To get around this problem we could pre-fix a set of $\phi$'s, one for
every member of a set of ranges
\{$\Gamma(U)\}$. We could then have $u$ pick out the precise $\phi$ to
use. This requires introduction of substantial additional structure
into the definition of devices however. (A somewhat related notion is
considered in Sec.~\ref{sec:sad}.) Another possible solution would be
along the lines of $\forall \phi : Y(U) \rightarrow \Gamma$, $\exists
x$ such that $X = x \Rightarrow \phi(Y) = \Gamma$''. But this returns
us to a definition of inference involving multiple functions relating
$Y$ and $\Gamma$.

All of these other difficulties also apply to the
definition above of joint inference involving multiple devices. In
particular, say we wish to use the same set of devices to jointly
infer function having different ranges from one another. Then we have
to specify something about how to map the joint conclusion of
the devices into an inference in any of those ranges. For example, if
the set of devices is \{$C_z : z \in {\mathbb{R}}$\} and $\Gamma(U)$
is non-numeric, we would need to specify something about how a joint
conclusion \{$Y_z(u)$\} gets mapped into that non-numeric space.

As a final possibility, we could stick with a single device and have
$Y(U) = {\mathbb{B}}$, but use some representation of $\Gamma(U)$ in
$X$ other than the unary representation implicit in Def. 3. For
example, we could require that for all binary representations $\phi$
of $\Gamma(U)$, for all bits $i$ in that representation, there is an
$x$ such that $X = x \Rightarrow Y = \phi_i(\Gamma)$. This would allow
smaller spaces $X(U)$ in general.  But it would still require
consideration of multiple functions relating $Y$ and $\Gamma$. It
would also raise the issue of how to encode the elements of
$\Gamma(U)$ as bits.

For simplicity, in the text we avoid these issues and restrict
attention to the original definitions.

\end{document}